\documentclass[twocolumn]{aastex61} 

\usepackage{csvsimple}
\usepackage{multirow,bigdelim}
\usepackage{amsmath}

\newcommand{\ms}{\mbox{m s$^{-1}$}}

\newcommand{\kms}{\mbox{km s$^{-1}$}}

\usepackage{colortbl}
\definecolor{tablegray}{rgb}{0.89, 0.89, 0.89}

\begin{document}

\title{Identifying Interstellar Object Impact Craters}
\correspondingauthor{Samuel H. C. Cabot}

\email{sam.cabot@yale.edu}

\author{Samuel H. C. Cabot}
\affil{Yale University, 52 Hillhouse, New Haven, CT 06511, USA}

\author{Gregory Laughlin}
\affil{Yale University, 52 Hillhouse, New Haven, CT 06511, USA}

\begin{abstract}

The discoveries of two Interstellar Objects (ISOs) in recent years has generated significant interest in constraining their physical properties and the mechanisms behind their formation. However, their ephemeral passages through our Solar System permitted only incomplete characterization. We investigate avenues for identifying craters that may have been produced by ISOs impacting terrestrial Solar System bodies, with particular attention towards the Moon. A distinctive feature of ISOs is their relatively high encounter velocity compared to asteroids and comets. Local stellar kinematics indicate that terrestrial Solar System bodies should have experienced of order unity ISO impacts exceeding 100 \kms. By running hydrodynamical simulations {for projectiles of different masses and impact velocities, up to 100 \kms}, we show how late-stage equivalence dictates that {transient} crater {dimensions are} {alone insufficient for inferring the projectile's velocity}. On the other hand, the melt volume within craters of a fixed diameter may be a potential route for identifying ISO craters, as faster impacts produce more melt. This method requires that the melt volume scales with the energy of the projectile, while crater diameter scales with the point-source limit (sub-energy). Given that there are probably only {a few} ISO craters in the Solar System {at best}, and that {transient} crater dimensions are not a distinguishing feature for impact velocities at least up to $100$ \kms, identification of an ISO crater proves a challenging task. Melt {volume} and high-pressure petrology may be {diagnostic} features once large volumes of material can be analyzed {\it in situ}. 

\end{abstract}


\section{Introduction} \label{sec:intro}

The discoveries of `Oumuamua \citep[from the Pan-STARRS survey, ][]{Meech2017} and Comet 2I/Borisov (by G. Borisov at the Crimean Astrophysical Observatory in 2019)\footnote{www.minorplanetcenter.net/mpec/K19/K19RA6.html} have prompted intensive study of the number density, composition, and origin of ISOs. Initial upper limits on number density were placed by \citet{Engelhardt2017}, based on simulated ISO populations and their detectability by modern surveys. However, the discovery of `Oumuamua yielded an estimate for similar objects of 0.2 au$^{-3}$ \citep{Do2018}. While Comet 2I/Borisov is very similar to Solar System comets \citep{Guzik2020}, `Oumuamua's oblong shape and lack of coma \citep{Meech2017}, along with its anomalous acceleration \citep{Micheli2018} has forced reconsideration of its makeup, including materials atypical of comets and asteroids \citep[e.g.][]{Rafikov2018, Fuglistaler2018, Desch2021}. Earlier identification with the Vera C. Rubin Observatory (LSST) or even {\it in situ} analyses \citep{Snodgrass2019} would drastically improve our understanding of ISOs; specifically their relationship to the galaxy-wide population of ejected planetesimals \citep{ Trilling2017}. 

The entry trajectory of `Oumuamua (at speed $v_\infty \simeq 26$ \kms; \citealt{Meech2017}) was similar to the local standard of rest (LSR) \citep{Francis2009}, consistent with expectations for ISOs. The difference between the median velocity of nearby stars (XHIP catalog; \citealt{Anderson2012}) and that of `Oumuamua's entry was only about $4.5$ \kms\ at $\sim 6^{\circ}$ \citep{Mamajek2017}. Nevertheless `Oumuamua was not comoving with any particular nearby system. While specific stars have been postulated as the origin, chaotic gravitational interactions make a precise back-tracing impossible. Unexpectedly perhaps, 2I/Borisov entered at $v_\infty \sim 32$ \kms\ at $\sim 75^{\circ}$ away from the solar apex \citep{Guzik2020}, its origin again speculative \citep{Dybczynski2019}. As pointed out by \citet{Do2018}, the detection volume of ISOs scales as $v_{\infty}^{-1}$, from multiplication between gravitational focusing from the Sun (the effective cross-section becomes $r_g^2 = r^2 [(v_{\rm esc}/v_\infty)^2 + 1]$) with the impingement rate. Therefore ISOs may be less efficiently detected if they encounter the Solar System at speeds substantially exceeding the Sun's escape velocity at $d\sim1\,{\rm au}$. The detectability of ISOs as a function of $v_\infty$ and impact parameter $b$ is quantified by \citet{Seligman2018}. ISOs with $v_\infty \gtrsim 10$ \kms\ must have $b \lesssim 5$ au if they are to be identified by LSST prior to periastron. Although `Oumuamua came serendipitously close to Earth ($r_p \simeq 0.25$ au, $b \simeq 0.85$ au), these calculations reveal the significant challenge of detecting additional ISOs. 

Motivated by an encouragingly high encounter rate of ISOs, up to $\sim 7$ per year that pass within $1\,{\rm au}$ of the Sun \citep{Eubanks2021}, we consider an alternative route to characterizing these enigmatic objects: identifying ISO impact craters on terrestrial Solar System bodies. {For example, molten and vaporized projectile matter may mix with impact-modified target rock (impactites) and impart tell-tale chemical signatures. More optimistically, some projecile material might survive in solid phase. A suite of standard chemical and isotopic analyses exist for characterizing meteorites and impact melts \citep{Tagle2006, Joy2016}, which could reveal the ISO's composition.}

{Before an {\it in situ} or retrieved sample analysis is possible, we need a high-fidelity method for screening ISO craters from asteroid and comet craters.
Crater morphology and high-pressure petrology may be differentiating traits; but this premise is significantly challenged by well-known degeneracies between crater and projectile properties \citep{Dienes1970, Holsapple1982}}. {Although, some} constraints have been achieved for especially renowned and well-studied craters. For example, \citet{Collins2020} used 3D simulations to link asymmetries in the Chicxulub crater to a steeply-inclined impact trajectory; although the observations are compatible with a modest range of angles and impact speeds. Using an atmospheric-entry fragmentation model, \citet{Melosh2005} {posited} that Meteor Crater was formed by a low-speed impact, which additionally explains an anomalously low melt volume. {However, this model was challenged by \citet{Artemieva2011} on the basis of little observed solid projectile ejecta.} As another example, \citet{Johnson2016} modeled formation of the Sputnik Planum basin and found consistency with a 220 km diameter projectile; however they assumed a 2 \kms\ speed typical for impacts on Pluto. {There is a considerable amount of literature surrounding each of these craters, which raises a number of other interpretations than those listed here \citep[e.g.][]{Artemieva2009, Denton2021}, and echoes the difficulties of inferring projectile properties from their craters.} {We note that impacts} in the Solar System virtually never exceed $\gtrsim$ 100 \kms, and hence these speeds are seldom modeled in the literature. Nevertheless, we will show they are not atypical for ISO impacts, and thus warrant further investigation --- this aspect is the main focus of our study.

{If ISO craters can be identified, then surviving ISO meteorites in and around the crater could be readily analyzed for metallic content, oxygen isotope fractionation, and elemental ratios (e.g. Fe/Mn) \citep{Joy2016}; however, if} ISOs are composed of highly volatile, exotic ice \citep{Seligman2020, Desch2021}, we may expect that they undergo near-complete vaporization upon impact, and suffer the same issues in chemical-based identification as comets do \citep{Tagle2006} (a small percentage of water content may survive comet impacts \citep{Svetsov2015}). {An ISO's composition could still be investigated if its material persists in the impact melt or vapor condensates. For example, }
\citet{Tagle2006} evaluate a few methods for projectile classification, involving relative concentrations of platinum group elements (PGEs), Ni and Cr, and isotopic ratios of Cr and Os. At present, `Oumuamua's composition is highly speculative, as is the composition of the general ISO population. {Any insight into their compositions can be directly tied to formation pathways (e.g. molecular cloud cores in the case of H$_2$, or cratered ice sheets in the case of N$_2$) as well as their abundance in the galaxy \citep[][and references therein]{Levine2021}}.

Our study is outlined as follows: In \S\ref{sec:vel} we review the impingement rate of ISOs and the expected velocity distribution based on local stellar kinematics. In \S\ref{sec:crater} we present hydrodynamical simulations representative of ISO impacts on terrestrial bodies. While certain aspects of these impacts are unconstrained (most notably the projectile composition) we use well-understood materials as proxies to obtain order-of-magnitude estimates of crater size and melt volume. We restrict analysis to transient craters for simplicity; although {collapse and viscous degradation may modify their shapes} \citep[][Chapter 8]{Melosh1989}. Specific attention is given to lunar cratering in light of soon-to-be realized exploration missions; however parts of our investigation extend to other terrestrial bodies such as Mars. The simulation results are subsequently compared to predictions from crater {scaling relationships}. We discuss additional scaling relations in \S\ref{sec:melt}, with a particular focus on how melt volume may be used to infer the impact velocity. Our results are summarized in \S\ref{sec:disc}.

\section{ISO Impact Velocities} \label{sec:vel}

It is important to determine the speed at which ISOs impact terrestrial bodies in the Solar System. {A significant component} is from $v_\infty$, the speed at which the ISO encounters the Solar System. About $40$ \kms\ is added in quadrature for ISOs that come within $1\,{\rm au}$ of the Sun). {ISO impacts on the Moon can reach velocities $\geq100$ \kms; these events are the focus of our study.} We review analytic expressions for the kinematics of stars in the solar neighborhood, as well as measurements of the velocity dispersion along each principal axis. Next, we independently analyze the kinematics of stars with full phase-space measurements provided in the recent {\it Gaia} data release. These velocities are combined with the estimated number density of ISOs to obtain the encounter rate as a function of ISO speed. 

\subsection{Local Stellar Kinematics: Theory} \label{subsec:veltheory}

 ISOs of icy composition are expected to have a kinematic distribution reflective of their origin systems. \citet{Binney2008} show velocities in the galactic disk are well-described by a Schwarzschild distribution
 
\begin{equation}
    f(\mathbf{v}) = S(L_z) \exp\Big[-\Big(\frac{v_R^2 + \gamma^2\widetilde{v}_\phi^2}{2\sigma_R^2(L_z)} + \frac{v_z^2 + 2\Phi_z(z, L_z)}{2\sigma_z^2(L_z)}\Big)\Big]\, ,
\end{equation}
for cylindrical velocity components $v_R$, $v_\phi$, and $v_z$ and their respective dispersions $\sigma_R$, $\sigma_\phi$, and $\sigma_z$ \citep{Dehnen1998, Nordstrom2004}. Angular momentum is denoted $L_z$. The term $\widetilde{v}_\phi \equiv v_\phi - v_c(R)\mathbf{\hat{e}}_\phi$ represents difference between the angular velocity component and the circular velocity at the star's galactic radius, $R$. The term $\gamma \equiv 2\Omega/\kappa$ arises from the guiding center approximation, where $\Omega$ is the circular frequency and $\kappa$ is the epicyclic frequency. The potential $\Phi_z(z, L_z)$ appears from an approximation to the third integral of motion. The exponential form follows from \citet{Shu1969}, and the leading term $S(L_z)$ depends on the surface density of stars. Under two approximations, first that the surface density follows an exponential disk, and second that the dispersions are relatively low compared to the circular speed (i.e. that the stars are of a ``cold" population), the solar neighborhood distribution follows a triaxial Gaussian model \citep{Schwarzchild1907},

\begin{equation}
    dn \propto \exp\Big[-\Big(\frac{v_R^2 + \gamma^2\widetilde{v}_\phi^2}{2\sigma_R^2} + \frac{v_z^2}{2\sigma_z^2}\Big)\Big].
\end{equation}
If one generalizes beyond the epicyclic approximation, which also assumed that $\sigma_\phi/\sigma_R = \kappa/2\Omega$, then the solar neighborhood distribution becomes

\begin{equation} \label{eqn:velcomp}
    f(\mathbf{v}){\rm d}^3\mathbf{v} = \frac{n_0 {\rm d}^3\mathbf{v}}{(2\pi)^{3/2}\sigma_R\sigma_\phi\sigma_z}
    \exp\Big[-\Big(\frac{v_R^2}{2\sigma_R^2} + \frac{\widetilde{v}_\phi^2}{2\sigma_\phi^2} + \frac{v_z^2}{2\sigma_z^2}\Big)\Big]\, ,
\end{equation}
where $n_0$ is the number of stars per unit volume \citep{Binney2008}. This equation is useful under the assumption that ISOs originate predominantly from nearby, Population I stars. As discussed in the following subsection, population studies provide excellent constraints on the dispersion along each principal axis. However, the distribution for speed $|v|$ is not well described by a Gaussian or Boltzmann distribution; a log-normal model provides a reasonable fit \citep{Eubanks2021}.

The impact rate of ISOs, $\Gamma = n_{\rm ISO}\sigma_p v_{\infty}$, depends on the number density of ISOs, the cross-sectional area of the target, and the relative velocity of the two bodies. A more detailed formulation is given by \citep{Lacki2021},

\begin{equation}
    \Gamma(\geq K_T) = \int_{0}^{\infty} \int_{2K_T/v_{\infty}}^{\infty} \sigma_p v_{\infty} f(v_{\infty}) \frac{dn_{\rm ISO}}{dm} dm d v_{\infty}\, ,
\end{equation}
which is the impact rate of ISOs with energy at least $K_T$. The mass distribution is probably well-described by a {power law}, which is often adopted for minor body populations in the Solar System. Order-of-magnitude estimates by \citet{Lacki2021} {yield an ISO impact rate of $6\times10^{-6}$ Gyr$^{-1}$ at Earth, restricted to projectiles with $\geq1$ YJ kinetic energy (roughly equivalent a $10^{15}$ kg projectile impacting at $45$ \kms)}. A one-dimensional, Maxwellian stellar velocity dispersion of 30 \kms\ was assumed, which is roughly the average of the three solar neighborhood dispersions measured by \citet{Anguiano2017}. {We investigate the local velocity dispersion in more detail in the next subsection}. Note the actual impact speed of the ISO is higher than the relative speed with the Solar System ($v_i > v_\infty$) due to extra energy gained by falling into the Sun's potential well, plus a small contribution {from} the target planet or {satellite}'s gravity, {notwithstanding atmospheric effects}. Also, the effective cross-section is modified by gravitational focusing.

{Our investigation hinges on the possibility that anomalously fast ISO impacts produce craters distinct from comet and asteroid impacts. Therefore, we review the distributions of impact speeds in the Solar System to determine a velocity threshold that effectively excludes comets and asteroids.} {Impacts at Earth, Venus, and Mercury commonly exceed 20 \kms\ \citep{lefeuvre2011}, with Mercury's distribution extending to 90 \kms. For the Earth/Moon system, impacts rarely occur at greater than 50 \kms\ \citep{lefeuvre2011}. The high-velocity tail mainly comprises long-period comets which may impact at speeds up to $\sim70$ \kms\ \citep[][]{Steel1998}. Cosmic velocities of ISOs, however, occasionally exceed 90 \kms\ (see below) and would therefore yield {impacts faster than expected for typical Solar System impactors} {(e.g. 100 \kms\ at Earth and up to 113 \kms\ at Mercury, taking into account the Sun's potential well and planet escape velocity)}. {Comet and asteroid impact velocities are generally lower for bodies at} larger semi-major axes. For example, the mean impact speed is 10.6 \kms\ for Mars \citep{lefeuvre2008} and 4.75 \kms\ for Vesta \citep{obrien2011}. The distribution of impacts vanishes past 40 \kms\ for Mars, and past 12 \kms\ for Vesta and Ceres \citep{obrien2011}. If craters could be linked to these impact speeds or higher, ISOs would be strong candidates for the associated projectile. Therefore, while this study is primarily concerned with impacts on the Moon, a larger range in impact speed could be associated with ISOs for craters on Mars and more distant terrestrial bodies.}

\subsection{Local Stellar Kinematics: Observed} \label{subsec:velobs}

\begin{figure}
    \centering
    \includegraphics[width=\linewidth]{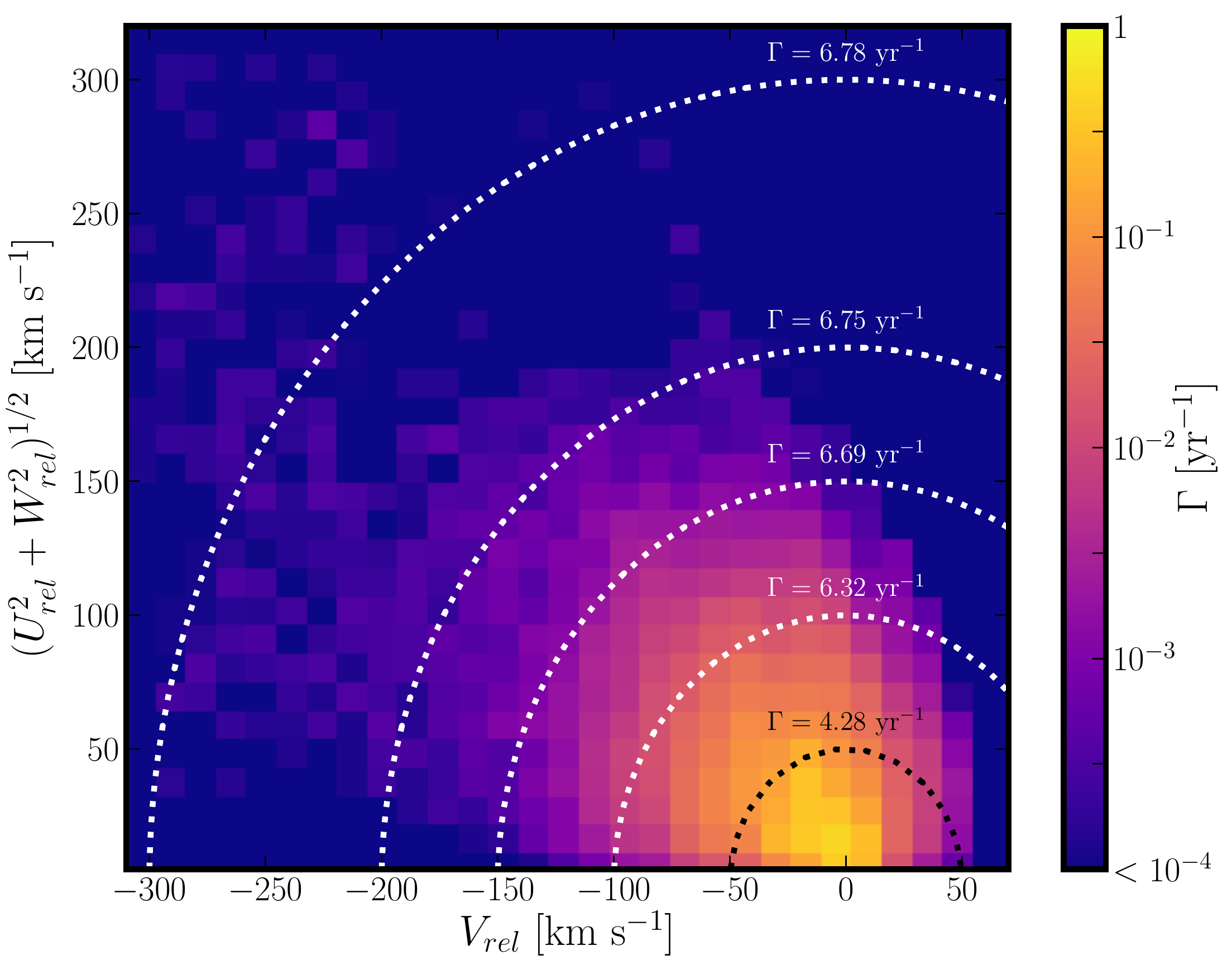}
    \caption{Toomre Diagram for {\it Gaia} EDR3 stars within the 200 pc of the Sun, which have full phase-space measurements. The origin represents the rest-frame of the Sun. Values have been normalized to depict the encounter rate of ISOs within 1 au of the Sun. Curves correspond to, from inner-most to outer-most, constant encounter speeds of 50 \kms, 100 \kms, 150 \kms, 200 \kms, and 300 \kms. Labels indicate the summed encounter rate of all ISOs with velocities enclosed by the curve.}
    \label{fig:rates}
\end{figure}

The proper motions of nearby stars are thoroughly measured thanks to large surveys. {\it Gaia}, for example, has provided a massive catalog of 7.2 million stars with complementary line-of-sight velocities \citep{GaiaCollaboration2018a}. After filtering, their main sample contained approximately 6.4 million sources with full phase-space measurements. The vast majority of stars within the sample lie near the origin in the classic Toomre diagram \citep{Sandage1987} depicting $V$ against $(U^2+W^2)^{1/2}$ offset by the solar LSR ($U$, $V$, and $W$ refer to radial, tangential, and vertical velocity components respectively). This figure is often used to depict distinct populations of stars \citep{Venn2004}. Iso-velocity contours in the Toomre Digram delineate transitions between stellar populations; for example, \citet{Nissen2004} define 80 \kms\ and 180 \kms\ as the boundaries confining thick-disk stars, where lower speeds correspond to the thin-disk stars. \citet{Venn2004} used the Toomre Diagram to dynamically classify stars into five categories (thin-disk, thick-disk, halo, high-velocity, and retrograde), and subsequently determine chemical properties of each population. 

A significant fraction of stars in the {\it Gaia} catalog have relative speeds exceeding $100$ \kms, but few lie in the Solar System's vicinity. For stars in the galactic mid-plane (extending $-200$ to $+200$ pc), velocity dispersions are of order $10-40$ \kms\ for the three components, with some variation in radial distance from the galactic center. Populations of stars that are a few kpc above and below the mid-plane exhbiit dispersions of up to $60-80$ \kms\ per component. Other survey studies also report the spatial dependence of velocity dispersion (generally increasing toward the galactic center, and away from the mid-plane) \citep[e.g.][]{Bond2010, Recio-Blanco2014}. Stellar properties such as metallicty and age are correlated with velocity dispersion \citep[e.g.][]{Stromgren1987, Nissen2004, Rojas-Arriagada2014}. Example dispersions considered by \citet{Binney2008} were based on the Geneva-Copenhagen survey \citep{Nordstrom2004} that observed F and G dwarfs. \citet{Nordstrom2004} presented age-dependent velocity dispersions. The youngest stars (within their 1 Myr age bin) had $\sigma_{\rm tot} \approx 30$ \kms, while the oldest stars (within their 10 Myr age bin) had $\sigma_{\rm tot} \approx 60$ \kms. In all bins, it was found $\sigma_U > \sigma_V > \sigma_W$. 

We analyzed the dynamics of ISOs originating within the local stellar neighborhood using {\it Gaia} EDR3 data \citep{Gaia2020}, in a similar fashion as \citet{Marchetti2020} and \citet{Eubanks2021}. The Sun's peculiar velocity was taken as (11.1, 12.24, 7.25) \kms\ \citep{Schonrich2010} relative to an LSR of (0, 235, 0) \kms. The dynamics of the closest stars are probably most representative of ISO velocities, so we included only stars within 200 pc of the Sun. The Toomre Diagram for the stellar sample is shown in Figure~\ref{fig:rates}, where velocity components are as measured in the Sun's rest-frame. Each bin is rescaled to reflect its contribution to the encounter rate of ISOs. This step is accomplished by first normalizing the sum over all bins to the ISO number density, $n_{\rm ISO} \sim$ 0.1 au$^{-3}$. This value is half the estimate of \citet{Do2018}, and is used as an upper limit by \citet{Eubanks2021} who appeal to the lack of recent detections. Each bin is multiplied by its speed $|v_\infty|=\sqrt{U_{rel}^2 + V_{rel}^2 + W_{rel}^2}$, a cross-section of 1 au$^2$, and a gravitational-focusing enhancement factor of $1 + (v_{\rm esc}/v_\infty)^2$, where $v_{\rm esc}$ is evaluated at 1 au. The results do not strongly depend on population volume since we normalize the distribution to reflect the total number density of ISOs. We find the total encounter rate of ISOs within 1 au of the Sun is about $6.80$ yr$^{-1}$. The majority arrive with $v_\infty < 100$ \kms, with a rate of 6.32 yr$^{-1}$. High-speed ISOs with $v_\infty > 100$ \kms\ arrive at 0.47 yr$^{-1}$, and $v_\infty > 200$ \kms\ arrive at 0.05 yr$^{-1}$. Our results are nearly the same as those of \citet{Eubanks2021}.

Interestingly, high-speed ISOs make a non-negligible contribution to the encounter rate, despite the vast majority of nearby stars having relative speeds of $\lesssim 100$ \kms\ (the peak of the distribution lies at around $40$ \kms). Multiplying by the ratio of the target's cross section to 1 au$^2$, we find Earth and the Moon experience $\sim 12$ and $\sim 0.9$ ISO impacts per Gyr, respectively. The objects most pertinent to this study, ISOs that impact the the Moon at speeds $v_i > 100$ \kms, have encounter speeds of $v_{\infty} > 90.6$ \kms, and a corresponding impact rate of $\sim 0.09$ per Gyr. Equivalently, there is a $31\%$ chance that the Moon experienced a high-speed ISO impact in the past 4 Gyr. Repeating the above analysis for Mars yields a high-speed impact rate of $\sim 0.29$ per Gyr. These results indicate that there should be of order unity high-speed ISO impact craters {on} the Moon and Mars.

{For most remaining terrestrial bodies, the chances of identifying an ISO crater based on the projectile's extreme speed appear slim. High-speed impacts of asteroids and comets are common at Mercury's orbit \citep{lefeuvre2011}; Venus experienced a recent cataclysmic resurfacing event \citep{Schaber1992}; and Earth's geological activity has largely erased ancient craters. The Galilean Moons Io and Europa seem unlikely candidates due to their small surface areas and young surface ages of $0.3-2.3$ Myr and 60 Myr, respectively \citep{Schenk2004}. Ganymede and Callisto on the other hand have surface ages of $\gtrsim 2$ Gyr and could be potential targets.}

\section{{Transient} Crater {Dimensions}} \label{sec:crater}

It is well known that crater dimensions are highly degenerate with projectile properties (e.g. velocity, radius, density, impact angle) \citep{Dienes1970, Holsapple1987}. We simulate impacts on the Moon in order to test whether degeneracies persist at the high-velocity tail of ISO impacts. {Our selection of target materials is a subset of those simulated by \citet{Prieur2017}, characteristic of the lunar regolith and upper megaregolith.} We then compare the results to theoretical expectations for crater diameter based on late-stage equivalence \citep{Dienes1970}.

\subsection{{Simulation Overview}}

We simulate impacts with the iSALE-Dellen 2D hydrocode \citep{Wunnemann2006}, which is based on the Simplified Arbitrary Lagrangian-Eulerian (SALE) program \citep{Amsden1980} designed for fluid flow at all speeds. SALE features a Lagrangian update step, an implicit update for time-advanced pressure and velocity fields, and finally an advective flux step for Eulerian simulations. Calculations are performed on a mesh in an Eulerian frame of reference to prevent highly distorted cells. Over the years, the program has seen new physics implemented, including an elasto-plastic constitutive model, fragmentation models, various equations of state (EoS), multiple materials, and new models of strength, porosity compaction, and dilatancy \citep{Melosh1992, Ivanov1997, Wunnemann2006, Collins2011, Collins2014}. Massless tracer particles moving within the mesh \citep{Pierazzo1997} record relevant Lagrangian fields. We adopt a resolution of 20 cells per projectile radius (CPPR) which has been demonstrated to be within $\sim10\%$ of convergent spall velocity \citep{Head2002}, peak shock pressure, and crater depth and diameter \citep{Pierazzo2008}. \citet{Barr2011} show that 20 CPPR underestimates melt volume by $\sim15\%$ in simulations of identical projectile and target materials. {For our impact configurations, we found $19\%$ and $22$\% lower melt volume in 20 CPPR simulations compared to 80 CPPR, for 30 \kms\ and 100 \kms\ impacts, respectively (Appendix~\ref{sec:appmelt}). Therefore we multiply melt volumes in our main analysis by a proportionate correction factor.} The timestep is limited by the Courant-Friedrichs-Levy (CFL) criterion, which demands higher temporal resolution for faster material speeds (and faster impact velocities). {We fixed the width of the high-resolution zone, which we found to overlay roughly the inner-half of the transient crater. This layout is sufficient for determining melt volume, and for determining the transient crater diameter (Appendix~\ref{sec:appscale}).} Three-dimensional simulations are occasionally used in the literature \citep[e.g.][]{Artemieva2004}. They are prohibitively expensive for our {investigation}, and unnecessary for exploring quantitative differences in crater profiles resulting from variable impact velocity. We restrict our analysis to head-on, azimuthally symmetric impacts. More information regarding computational methods for impact simulations is discussed by \citet{Collins2012} and references therein. 

We focus attention to impacts on the Moon that produce {simple} craters. {Both 2I/Borisov and `Oumuamua have effective radii upper bounded by a few hundred meters, and the radii were more likely $\lesssim$ 100m \citep[e.g.][]{Jewitt2019}, which is insufficient to yield complex craters on the Moon.} We assume a target comprised of basalt and projectile of water ice. We acknowledge that `Oumuamua was likely not composed of water ice, and that 2I/Borisov was depleted in H$_2$O. The typical composition of ISOs remains debated, however, and all recent hypotheses have specific, production-limiting aspects \citep{Levine2021}. Nevertheless, `Oumuamua's anomalous acceleration probably implies a significant volatile component, either in the form of common ices (e.g. H$_2$O, CO) or exotic ices (e.g. H$_2$, N$_2$, CH$_4$) or a combination of both. We restrict analysis to a water ice projectile since: (1) the purpose of this study is to investigate whether extremely fast ISO impacts are discernible from those of comets and asteroids, and the main parameter of interest is impact speed; (2) the bulk properties of exotic ices are poorly constrained; and (3) many material properties of H$_2$O ice are within the same order of magnitude of those of other ices. {Nevertheless, an exotic ice projectile composition could affect the crater in a variety of ways. For example, extremely low density H$_2$ ($\rho \sim 0.08$ g cm$^{-3}$) would produce a crater of lower volume, owing to a shallower {penetration} depth $d_b \propto \rho_p^{0.5}$ \citep{Birkhoff1948}. Impacts on Mars are not thoroughly investigated here, but would warrant consideration of the planet's thin atmosphere. Exotic ice projectiles would fragment and thus modify the crater morphology \citep{Schultz1985}. Highly volatile ices may also experience increased ablation at lower velocities, reducing the projectile's mass.} 

\subsection{{Simulated Target and Projectile Properties}}

{Material specifications for our simulations are described as follows, and are also listed in Table~\ref{tab:params}. They are primarily based on parameters used by \citet{Prieur2017} for basalt and by \citet{Johnson2016} for water ice. Material strength is set by a Drucker-Prager model, which is most appropriate for granular targets. Required parameters include cohesion $Y_0$, coefficient of friction $f$, and limiting strength at high pressure $Y_{\rm LIM}$. The $\epsilon$-$\alpha$ compaction porosity model \citep{Wunnemann2006, Collins2011} is adopted for the target, but neglected for the projectile. The required parameters are initial distension $\alpha_0\equiv 1/(1-\Phi)$ (for porosity $\Phi$), elastic volumetric strain threshold $\epsilon_{e0}$, transition distension $\alpha_X$, compaction rate parameter $\kappa$, and sound speed ratio $\chi$. Tensile failure remains off since the target is already assumed damaged under the strength model. Acoustic fluidization is neglected since our simulations only concern simple craters. Dilatancy is also neglected since it has very small effect on transient crater dimensions \citep{Collins2014}. Low density weakening (a polynomial function of density) and thermal softening \citep{Ohnaka1995} are enabled.}

{We proceed to simplify our model by fixing several of the above variables. Porosity parameters are set to $\epsilon_{e0}=0$, $\alpha_X=1$, and $\kappa=0.98$ \citep{Collins2011}, as well as $\chi=1$ \citep{Wunnemann2006, Prieur2017}. We fix $f=0.6$ which is reasonable for sand-like materials. This value was used in early basalt target modeling \citep{Pierazzo2005}, the majority of models in a multi-layer lunar cratering study \citep{Prieur2018}, and in more recent impact studies involving basalt targets \citep[e.g.][]{Bowling2020}. Limiting strength $Y_{\rm LIM}$ has marginal effect on crater scaling parameters \citep{Prieur2017}; it is fixed to 1 GPa in our simulations. For the water ice projectile we fix $Y_0 = 0.01$ MPa, $f=0.55$, and $Y_{\rm LIM} = 147$ MPa \citep{Johnson2016}}. 

{The remaining material parameters are target $Y_0$ and $\Phi$. The lunar crust has an average porosity $\Phi=12\%$ extending a few km deep \citep{Wieczorek2013}, with variations between $4-21\%$. We perform simulations for three representative values of porosity: $\Phi=0,12,20\%$. We also consider two possibilities for cohesion: $Y_0=5$ Pa and $Y_0 = 10$ MPa. The former is representative of granular targets with negligible cohesion \citep[identical to][]{Prieur2017}, while the latter is representative of more competent targets. A cohesion of $10$ MPa is the highest cohesion considered by \citet{Prieur2017}, and may overestimate the actual cohesion in the heavily fractured and brecciated upper-megaregolith; but we adopt $10$ MPa for greater contrast against the nearly cohesionless scenario. We use an ANEOS equation of state (EoS) for the basalt and a Tillotson EoS for water ice (parameter values are listed in Table~\ref{tab:params}).}

{For each target material combination ($Y_0$, $\Phi$), we simulated nine impacts spanning projectile diameters $L=40, 80, 160$ m and velocities $v_i = 10, 30, 100$ \kms. A total of 54 simulations were performed.}\footnote{However, the \{$L=40$ m, $v_i = 100$ \kms, $\Phi=12\%$, $Y_0 = 5$ MPa\} simulation was not numerically stable, and is excluded from further analysis.}

\setlength{\tabcolsep}{1.0pt}
\renewcommand{\arraystretch}{1.0}
\begin{table}
\centering
\begin{tabular}{l r r} 
 \hline
 iSALE Material Parameter & Target & Projectile \\
 \hline\hline
Material                        & Basalt       & Ice\\
EOS type                        & ANEOS        & Tillotson\\
Poisson ratio                   & 0.25$^a$         & 0.33$^b$\\
Thermal softening constant      & 1.2$^a$          & 1.84$^b$\\
Melt temperature (K)            & 1360$^a$         & 273$^b$\\
Simon $a$ parameter (Pa)        & 4.5$\times10^{9,a}$ & 6.0$\times10^{9,c}$ \\
Simon $c$ parameter             & 3.0$^a$          & 3.0$^c$ \\
$^*$Cohesion (damaged) (Pa)     & (5, 1.0$\times10^7$)  & 1.0$\times10^{4,b}$ \\
Friction coeff. (damaged)       & 0.6$^a$          & 0.55$^b$ \\
Limiting strength (Pa)          & 1.0$\times10^{9,a}$  & 1.47$\times10^{8,b}$ \\
$^*$Initial Porosity ($\%$)          & (0, 12, 20)  & - \\
Elastic threshold                & 0.0$^a$              & - \\
Transition distension           & 1.0$^a$              & - \\
Compaction rate parameter       & 0.98$^a$             & - \\
Bulk sound speed ratio          & 1.0$^a$              & - \\
 \hline
Tillotson EoS Parameter (Ice)               & & Value \\
 \hline
 \hline
Reference density (g cm$^{-3}$)                & & 0.91$^c$\\
Spec. heat capacity (J kg$^{-1}$ K$^{-1}$)  & & 2.05$\times10^{3,c}$\\
Bulk modulus (Pa)                              & & 9.8$\times10^{9,c}$\\
Tillotson B constant (Pa)                      & & 6.5$\times10^{9,c}$\\
Tillotson E$_0$ constant (J kg$^{-1}$)         & & 1.0$\times10^{7,c}$\\
Tillotson a constant                           & & 0.3$^c$\\
Tillotson b constant                           & & 0.1$^c$\\
Tillotson $\alpha$ constant                    & & 10.0$^c$\\
Tillotson $\beta$ constant                     & & 5.0$^c$\\
SIE incipient vaporisation (J kg$^{-1}$)       & & 7.73$\times10^{5,c}$\\
SIE complete vaporisation (J kg$^{-1}$)        & & 3.04$\times10^{6,c}$\\
 \hline
\end{tabular}
\caption{{Table of parameters used in our hydrodynamical simulations. The basalt ANEOS is from \citet{Pierazzo2005}. An asterisk ($^*$) denotes parameters varied in our simulations. All fixed parameters include a reference: $^a$\citep{Prieur2017}, $^b$\citep{Johnson2016}, $^c$(parameter included in the iSALE-Dellen 2D distribution). The ice Tillotson EoS parameters are listed in the bottom section of the table. SIE $\equiv$ specific internal energy.} }
\label{tab:params}
\end{table}

\subsection{Expectations from Late-Stage Equivalence}

Late-stage equivalence, established by \citet{Dienes1970}, indicates that information surrounding the projectile is lost in the late stages of crater formation. Indeed, \citet{Holsapple1987} show that volume of the resultant crater, for a fixed combination of impactor and target materials, can be estimated by treating the projectile as a point-source characterized by coupling parameter 

\begin{equation} \label{eqn:couple}
    C = C(L, {v_i}, \rho_p) = Lv_i^{\mu}\rho_p^{\nu}.
\end{equation} 
The {power law} form follows from the requirements that $C$ remains finite as {projectile diameter} $L \xrightarrow[]{} 0$, and that $C$ must have fixed dimensionality. The convention adopted by \citet{Holsapple1987} is that $C$ has unity length units.
Impacts with equal $C$ produce transient craters with equal volumes. {\citet{Housen2011} review constraints on $\mu$ and $\nu$ from various past experiments. They indicate $\mu \sim 0.55$ for impacts into competent, non-porous rocks, which represents scaling in between momentum and energy dependence. Dry soils have $\mu \sim 0.41$, and highly porous materials are expected to have $\mu < 0.4$. Also, $\nu = 0.4$ has been shown to hold for a variety of materials, even when projectile and target bulk densities differ significantly.}

Using Pi-group scaling \citep{Buckingham1914}, one may choose dimensionless parameters:

\begin{equation} \label{eqn:pi1}
    \pi_D \equiv D_{tr}\Big( \frac{\rho_t}{m} \Big)^{1/3}
\end{equation}
\begin{equation} \label{eqn:pi2}
    \pi_2 \equiv \Big(\frac{4\pi}{3}\Big)^{1/3}\frac{gL}{v_i^2}
\end{equation}
\begin{equation} \label{eqn:pi3}
    \pi_3 \equiv \frac{Y}{{\rho_t} v_i^2}
\end{equation}
\begin{equation} \label{eqn:pi4}
    \pi_4 \equiv \frac{\rho_t}{\rho_p}
\end{equation}
\citep{Holsapple1982}, where $D_{tr}$ is the diameter of the transient crater, and $m$ is the projectile mass. {The material strength $Y$ is not precisely defined, but relates to cohesion and tensile strength.} The transient crater geometry is often used in studies of scaling relations, since it is not dependent on modification (there is also a slight distinction between rim-to-rim dimensions and `apparent' dimensions which are measured with respect to the pre-impact baseline). A properly chosen dimensionless functional relationship $\pi_D = F(\pi_2, \pi_3, \pi_4)$ often serves as a reasonable approximation for crater geometry. {\citet{Holsapple1982} provide a general scaling relation}

{
\begin{equation} \label{eqn:piD1}
    \pi_D = K_1\Big[\pi_2\pi_4^{\frac{2+\mu-6\nu}{-3\mu}} + \Big( K_2\pi_3\pi_4^{\frac{2-6\nu}{-3\mu}} \Big)^{\frac{2+\mu}{2}}  \Big]^{\frac{-\mu}{2+\mu}},
\end{equation}
for empirically determined scaling constants $K_1$ and $K_2$ (it is more useful to measure $K_2 Y$, rather than both individual terms).} Energy and momentum scaling {correspond to} $\mu = 2/3$ and $\mu = 1/3$, respectively, which is readily seen by taking the cube of $C$.
{Two regimes are apparent in the above equation: gravity-dominated craters (large $\pi_2$ term), and strength-dominated craters (large $\pi_3$ term). The former regime is appropriate for craters in the fine-grained lunar regolith, which is of order $10$ m deep \citep{McKay1991}. The lunar megaregolith consists of coarser-grained and heavily-brecciated material and extends tens of km deep, and cohesion likely factors into crater formation in this layer. We can use Pi-group scaling to predict which regime our simulations fall into. The transition between regimes occurs roughly when $\pi_2 = (K_2\pi_3)^{(2+\mu)/2}$, or equivalently $(K_2 Y / \rho_t v_i^2)^{1.25} \approx 1.6 g L/v_i^2$, assuming a typical $\mu = 0.5$. Approximating $K_2 Y \approx 20.9Y_0$ \citep{Prieur2017} and solving for $Y_0$, we can find the transition cohesive strength. For example, a 160 m diameter projectile striking at 100 \kms\ yields $\sim 2$ MPa. Therefore, our simulations of $Y_0 = 10$ MPa targets are in the strength-dominated regime, whereas those with $Y_0 = 5$ Pa targets are in the gravity-dominated regime. The same holds for other considered projectile diameters and velocities.}

\subsection{Results}

\begin{figure}
    \centering
    \includegraphics[width=\linewidth]{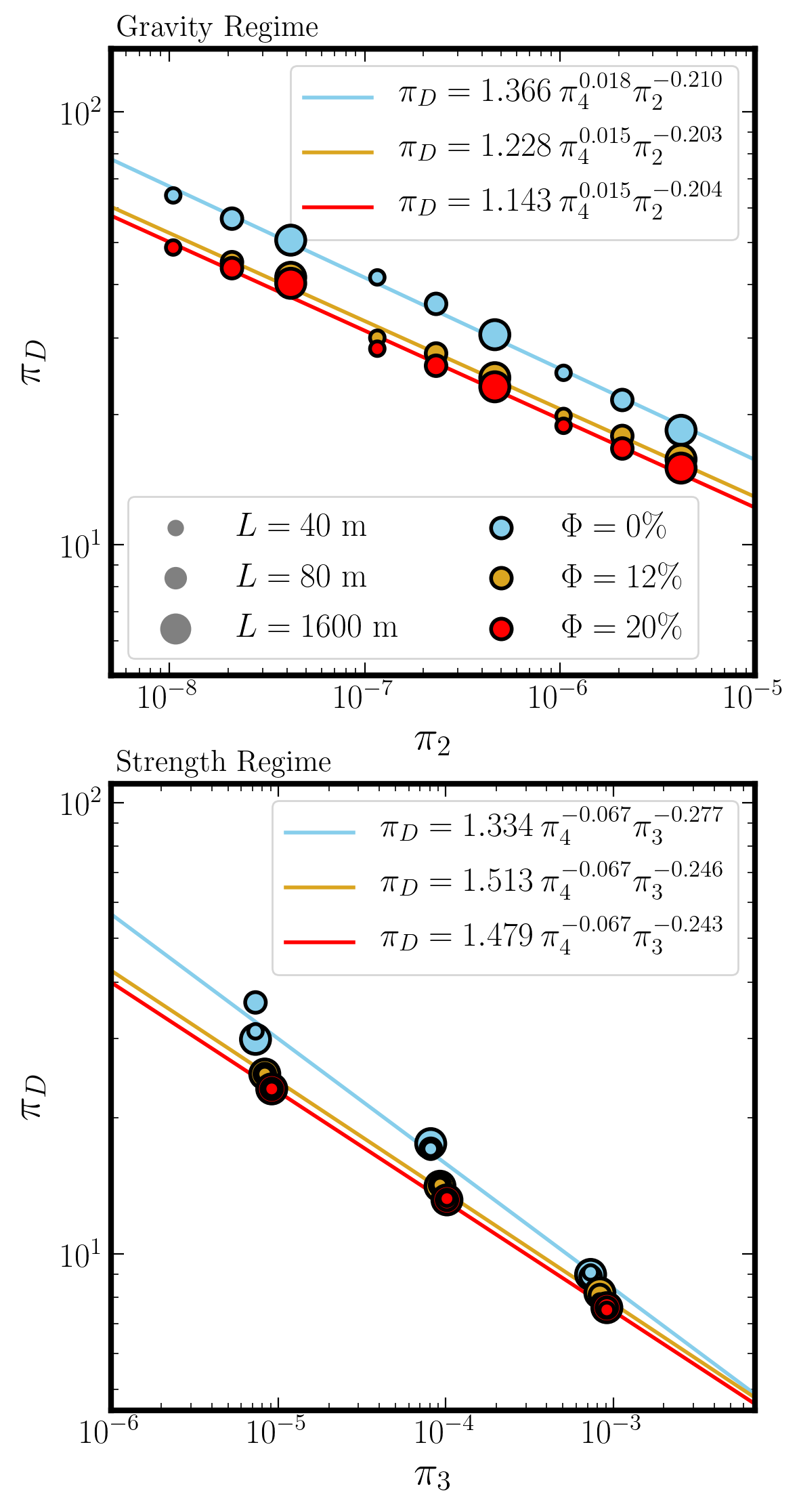}
    \caption{{Dimensionless scaling parameter outcomes for the iSALE simulations. Results for gravity-dominated craters are shown in the top panel, whereas those for strength-dominated craters are in the bottom panel. Colors denote three different target porosities, and the sizes of data points are proportional to the projectile diameter. The best-fit power law equations are denoted in the top-right corner of each panel.}}
    \label{fig:fits}
\end{figure}

{The simulations closely follow trends consistent with late-stage equivalence --- power law functions of the dimensionless Pi-group scaling parameters. The results are shown in Figure~\ref{fig:fits} for both the gravity- and strength-dominated regimes. In the former case we fit a power law between $\pi_D$ and $\pi_2$, and in the latter case we fit a power law between $\pi_D$ and $\pi_3$; subsequently, we solve for $\mu$. Outcomes for the three target porosities/distentions were fit separately, since they represent distinct target materials. In the gravity-dominated regime, we find $\mu = (0.533, 0.510, 0.514)$ for $\Phi=(0\%, 12\%, 20\%)$ scenarios. In the strength-dominated regime, we find $\mu = (0.554, 0.493, 0.486)$. Across all fits, the maximum deviation of $\pi_D$ from a power law fit is $7\%$. 
Discrepancies are addressed in \S\ref{sec:disc}, but the overall conformity of the dimensionless scaling parameters to a power law relationship confirms sub-energy scaling of crater diameter for projectile speeds up to 100 \kms}.

\begin{figure*}
    \centering
    \includegraphics[width=\linewidth]{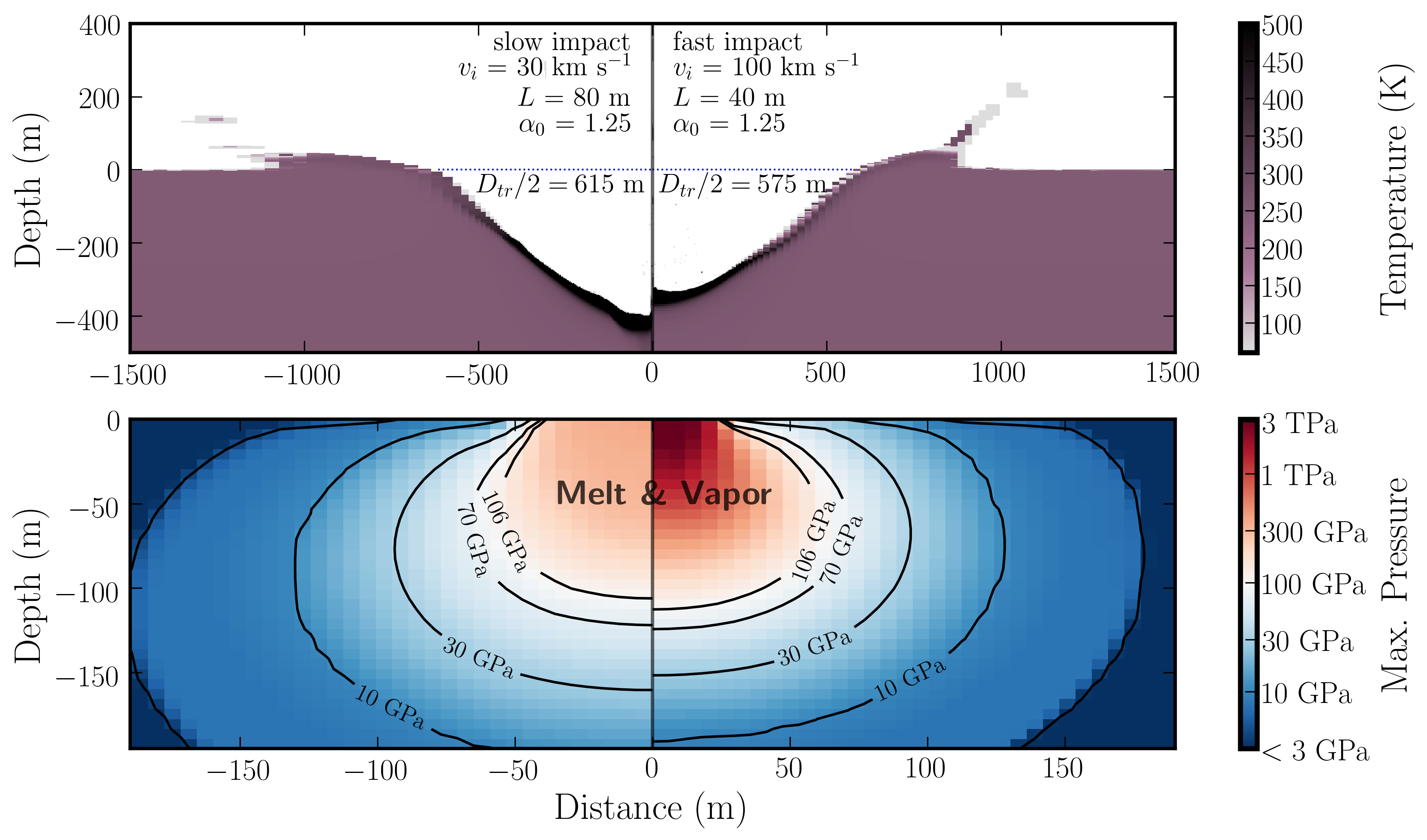}
    \caption{Two {example} simulations of ice projectiles {impacting} into a basalt target, {shown to highlight similarities in transient craters under different impact conditions}. Left panels correspond to a slow impact ({30} \kms) and large projectile diameter ({80} m), and the right panels correspond to a fast impact ({100} \kms) and small projectile diameter ({40} m). {The simulated targets have identical initial target distensions ($\alpha_0$) and negligible cohesive strength.} Top panels depict the transient crater profiles. {The reported {transient radii} ($D_{tr}/2$) are measured relative to the pre-impact surface.} The bottom panels depict the peak shock pressures experienced by tracer particles, which were embedded in the {high-}resolution zone. The contours are derived from the initial positions tracer particles. The critical shock pressure for {complete} melting is $P_c = 106$ GPa. Note the different abscissa and ordinate scales between the top and bottom panels.}
    \label{fig:profiles}
\end{figure*}

{In order to highlight the difficulty of inferring projectile characteristics from transient crater diameter alone, snapshots of two simulations are shown in Figure~\ref{fig:profiles}. One represents a slow, large projectile whereas the other represents a fast, small projectile impacting the same target material. Both simulations involved negligible target cohesive strength. Their transient diameters differ by $\sim 1\%$.} The diameters are `apparent' (i.e. measured at the level of the pre-impact surface. {There are slight differences in their (transient) profiles and depths; however, we do not explore these aspects in detail, since they will largely change in the subsequent modification stage.} Figure~\ref{fig:profiles} also shows contours of peak shock pressure, which may be used to infer melt volume. This point is investigated in \S\ref{sec:melt}.

\renewcommand{\arraystretch}{0.70}
\setlength{\tabcolsep}{14.0pt}

\begin{table*}
\centering
\footnotesize
\begin{tabular}{l l l l l l | r r} 
 \hline
 & & & & & & & \\
 $L$ (m) & $v_i$ (km s$^{-1}$) & $Y_0$ (Pa) & $\rho_t^{\rm ref}$ (g cm$^{-3}$) & $\rho_p$ (g cm$^{-3}$) & $\Phi$ ($\%$) & $D_{tr}$ (km) & $V_M^*$ (m$^3$ $\times 10^5$) \\
 & & & & & & & \\
 \hline\hline
 & & & & & & & \\
40 & 10 & 5 & 2.86 & 0.91 & 0 & 0.55 & 0 \\
40 & 10 & 5 & 2.86 & 0.91 & 12 & 0.46 & 0 \\
40 & 10 & 5 & 2.86 & 0.91 & 20 & 0.45 & 0 \\
40 & 10 & $1 \times 10^{7}$ & 2.86 & 0.91 & 0 & 0.20 & 0 \\
40 & 10 & $1 \times 10^{7}$ & 2.86 & 0.91 & 12 & 0.18 & 0 \\
40 & 10 & $1 \times 10^{7}$ & 2.86 & 0.91 & 20 & 0.01 & 0 \\
40 & 30 & 5 & 2.86 & 0.91 & 0 & 0.91 & 1.95 \\
40 & 30 & 5 & 2.86 & 0.91 & 12 & 0.69 & 1.76 \\
40 & 30 & 5 & 2.86 & 0.91 & 20 & 0.67 & 1.66 \\
40 & 30 & $1 \times 10^{7}$ & 2.86 & 0.91 & 0 & 0.38 & 1.95 \\
40 & 30 & $1 \times 10^{7}$ & 2.86 & 0.91 & 12 & 0.33 & 1.76 \\
40 & 30 & $1 \times 10^{7}$ & 2.86 & 0.91 & 20 & 0.31 & 1.66 \\
40 & 100 & 5 & 2.86 & 0.91 & 0 & 1.41 & 13.61 \\
40 & 100 & 5 & 2.86 & 0.91 & 20 & 1.15 & 11.45 \\
40 & 100 & $1 \times 10^{7}$ & 2.86 & 0.91 & 0 & 0.95 & 13.59 \\
40 & 100 & $1 \times 10^{7}$ & 2.86 & 0.91 & 12 & 0.57 & 12.13 \\
40 & 100 & $1 \times 10^{7}$ & 2.86 & 0.91 & 20 & 0.55 & 11.44 \\
80 & 10 & 5 & 2.86 & 0.91 & 0 & 0.95 & 0 \\
80 & 10 & 5 & 2.86 & 0.91 & 12 & 0.45 & 0 \\
80 & 10 & 5 & 2.86 & 0.91 & 20 & 0.79 & 0 \\
80 & 10 & $1 \times 10^{7}$ & 2.86 & 0.91 & 0 & 0.39 & 0 \\
80 & 10 & $1 \times 10^{7}$ & 2.86 & 0.91 & 12 & 0.37 & 0 \\
80 & 10 & $1 \times 10^{7}$ & 2.86 & 0.91 & 20 & 0.36 & 0 \\
80 & 30 & 5 & 2.86 & 0.91 & 0 & 1.59 & 15.55 \\
80 & 30 & 5 & 2.86 & 0.91 & 12 & 1.27 & 14.08 \\
80 & 30 & 5 & 2.86 & 0.91 & 20 & 1.23 & 13.29 \\
80 & 30 & $1 \times 10^{7}$ & 2.86 & 0.91 & 0 & 0.75 & 15.63 \\
80 & 30 & $1 \times 10^{7}$ & 2.86 & 0.91 & 12 & 0.65 & 14.07 \\
80 & 30 & $1 \times 10^{7}$ & 2.86 & 0.91 & 20 & 0.62 & 13.31 \\
80 & 100 & 5 & 2.86 & 0.91 & 0 & 2.50 & 108.87 \\
80 & 100 & 5 & 2.86 & 0.91 & 12 & 2.07 & 96.84 \\
80 & 100 & 5 & 2.86 & 0.91 & 20 & 2.07 & 91.08 \\
80 & 100 & $1 \times 10^{7}$ & 2.86 & 0.91 & 0 & 1.59 & 108.78 \\
80 & 100 & $1 \times 10^{7}$ & 2.86 & 0.91 & 12 & 1.15 & 96.78 \\
80 & 100 & $1 \times 10^{7}$ & 2.86 & 0.91 & 20 & 1.10 & 91.35 \\
160 & 10 & 5 & 2.86 & 0.91 & 0 & 1.62 & 0 \\
160 & 10 & 5 & 2.86 & 0.91 & 12 & 1.45 & 0 \\
160 & 10 & 5 & 2.86 & 0.91 & 20 & 1.43 & 0 \\
160 & 10 & $1 \times 10^{7}$ & 2.86 & 0.91 & 0 & 0.79 & 0 \\
160 & 10 & $1 \times 10^{7}$ & 2.86 & 0.91 & 12 & 0.75 & 0 \\
160 & 10 & $1 \times 10^{7}$ & 2.86 & 0.91 & 20 & 0.72 & 0 \\
160 & 30 & 5 & 2.86 & 0.91 & 0 & 2.69 & 124.35 \\
160 & 30 & 5 & 2.86 & 0.91 & 12 & 2.24 & 112.85 \\
160 & 30 & 5 & 2.86 & 0.91 & 20 & 2.20 & 106.49 \\
160 & 30 & $1 \times 10^{7}$ & 2.86 & 0.91 & 0 & 1.54 & 124.73 \\
160 & 30 & $1 \times 10^{7}$ & 2.86 & 0.91 & 12 & 1.30 & 112.44 \\
160 & 30 & $1 \times 10^{7}$ & 2.86 & 0.91 & 20 & 1.25 & 106.43 \\
160 & 100 & 5 & 2.86 & 0.91 & 0 & 4.45 & 869.22 \\
160 & 100 & 5 & 2.86 & 0.91 & 12 & 3.81 & 773.89 \\
160 & 100 & 5 & 2.86 & 0.91 & 20 & 3.81 & 729.39 \\
160 & 100 & $1 \times 10^{7}$ & 2.86 & 0.91 & 0 & 3.49 & 870.34 \\
160 & 100 & $1 \times 10^{7}$ & 2.86 & 0.91 & 12 & 3.01 & 772.02 \\
160 & 100 & $1 \times 10^{7}$ & 2.86 & 0.91 & 20 & 2.20 & 732.64 \\
 & & & & & & & \\
\hline
\end{tabular}
\caption{{Summary of hydrodynamic simulations. {Parameters left of the divider denote, from left to right, projectile diameter, impact speed, target cohesive strength, target reference density (i.e. notwithstanding porosity), projectile density, and porosity. Measured quantities right of the divider are transient crater diameter and melt volume. $^*$Reported melt volume is higher than simulation output owing to a correction ($23-28\%$ increase) that accounts for spatial resolution (Appendix~\ref{sec:appmelt}).}}}
\label{tab:simres}
\end{table*}

\section{Impact Melt Volume} \label{sec:melt}

As discussed above, there are well-known degeneracies between projectile mass, velocity, and impact angle in forming a crater. However, combinations of scaling relationships offer an opportunity to isolate variables of interest. Melt production {is of particular interest because it generally does not scale according to the point-source limit \citep{Pierazzo1997}}. {As a relatively recent example, \citet{Silber2018} simulated impacts of dunite projectiles into the Moon with $v_i$ ranging from $6$ \kms\ to $20$ \kms. They found a two order-of-magnitude difference in melt volume in craters with equal diameter, which shows the potential of using crater observables to deduce impact velocity.} In our investigation of scaling relations, we restrict analysis to vertical impacts and neglect dependence on impact angle.

\subsection{Numerical Simulations of Melt Production}

Melt volume in numerical simulations may be estimated by recording the peak shock pressure experienced by Lagrangian tracer particles \citep[e.g.][]{Wunnemann2008}. Plastic deformation from the shock wave irreversibly heats the target. If the target is shocked to a sufficient pressure, then it lies above the melt temperature following isentropic release from the rarefaction wave. A critical shock pressure {for complete melting} $P_c = 106$ GPa is adopted for basalt \citep{Quintana2015}.

The lower panels of Figure~\ref{fig:profiles} show the peak shock pressures experienced in {two representative examples of} our hydro simulations as a function of initial location in the target. The faster impact generates significantly higher peak pressures overall. {In our presentation of results,} we combine melt and vapor into a single `melt' volume {wherever peak shock pressures exceed $P_c$}. {Per Appendix~\ref{sec:appmelt}, all melt volumes were scaled by a correction factor to account for the simulation resolution of 20 CPPR. Melt volumes from all simulations are listed in the last column of Table~\ref{tab:simres}. Some immediately recognizable trends include: melt volume spans approximately three orders of magnitude, where the greatest melt volumes arise from the largest, fastest projectiles; only 30 \kms\ and 100 \kms\ impacts generated non-trivial melt volumes; holding other variables constant, target cohesion affects melt volume at a $\lesssim 10\%$ level in our simulations; and zero porosity yields $\sim20\%$ greater melt volume than the most porous materials explored.} 

{Can enhanced melt volumes assist in identifying the highest-speed impacts? Presence of significant basaltic melt can immediately rule out $\leq 10$ \kms\ impacts. However, at a constant $D_{tr}$, melt volume differences between $30$ \kms\ impacts and $100$ \kms\ impacts are more subtle. For example, $100$ \kms\ impacts of $40$ m projectiles produce similar $D_{tr}$ and melt volume as $30$ \kms\ impacts of $80$ m projectiles. Figure~\ref{fig:profiles} depicts this comparison for two example simulations. Note, the larger, slower projectile does yield a larger transient crater diameter and less melt; however, if actual lunar craters exhibited these properties, the differences between these two cases are probably too small to differentiate. Therefore, melt volume may be a important metric for filtering out low-speed asteroid impacts, but is less useful at the high-speed tail of the impact speed distribution, for these specific combinations of projectile and target materials. We proceed to place the simulation results in the context of established scaling relations}.

\subsection{Scaling Relations of Crater Dimensions and Melt Volume}

\citet{Pierazzo1997} performed hydrocode simulations of impacts with various materials, and fit a {power law} of the form

\begin{equation} \label{eqn:meltmu}
    \log\Big(\frac{V_M}{V_{p}}\Big) = a + \frac{3}{2}\mu'\log\Big(\frac{v_i^2}{E_M}\Big),
\end{equation}
a relation originally considered by \citet{Bjorkman1987}. {In the above, $a=\log k$, where $k$ is constant of proportionality that arises because the equation is based on dimensional analysis}, $V_p$ denotes the projectile volume, {$V_M$ denotes melt volume}, and $\mu'$ is a scaling constant. $E_M$ is the {specific} energy of melting. Values of $E_M$ for {several} materials of interest are listed by \citet{Bjorkman1987} and \citet{Pierazzo1997}, {as well as \citet{Quintana2015} for basalt}. 

{In general $\mu \neq \mu'$ because transient crater diameter scales according to the point-source limit, whereas melt volume does not. Indeed,}
{\citet{Okeefe1977} and \citet{Pierazzo1997} suggested $\mu'$ is consistent with $2/3$ (energy scaling). More recent works \citep{Barr2011, Quintana2015} reaffirm energy scaling for melt numbers $v_i^2/E_M \gtrsim 30$.} {Meanwhile, $\mu< 2/3$ for many materials of interest \citep{Schmidt1987}.} {In an ideal situation and holding all other variables constant,} combined measurements of crater diameter and melt volume {can in theory} break the degeneracy between projectile mass and velocity.  {This premise is elaborated upon in Appendix~\ref{sec:appmeltvol} where we derive equations for melt volume as a function of impact velocity and transient crater diameter, and demonstrate
}

{
\begin{equation} \label{eqn:meltprop}
    V_M \propto D_{\rm tr}^x v_i^{3(\mu'-\mu)},
\end{equation}
for sufficiently fast impacts. The constant of proportionality depends on the materials involved. In the strength-dominated regime, $x=3$, and in the gravity-dominated regime, $x=(6+3\mu)/2$. This relationship is independent of $m$ and $L$, so one may in principle solve for $v_i$ from two crater measurements. {In practice, impact angle, target lithology, and the variable composition of ISOs and other projectiles add degeneracies which would significantly complicate efforts to find an ISO crater. Additionally, long-term modification processes may alter crater morphology and make inferences of $D_{tr}$ less accurate. However, our exploration is designed to gauge the baseline feasibility of crater identification using these two observables, which may serve as a starting point for more sophisticated models that employ other sources of data (e.g. those discussed in \S\ref{sec:disc}).}
}

{As follows, we make a theoretical quantification of melt volume, and draw comparison to our hydro simulations.} {The analysis requires determining the constant of proportionality in Equation~\ref{eqn:meltprop}, which depends non-trivially on material properties including coefficient of friction, porosity, and cohesive strength, in addition to impact angle \citep{Schmidt1987, Elbeshausen2009, Prieur2017}. We take Equations~\ref{eqn:fullscale} $\&$ \ref{eqn:fullscale2} to analytically describe melt production {in the gravity-dominated and strength-dominated regimes, respectively, and rearrange to obtain} a function for impact velocity.} We adopt a melt energy $E_M = 8.7\times10^6$ J kg$^{-1}$ for the basalt target \citep{Quintana2015} with density $\rho_t = 2.86$ g cm$^{-3}$ (modified accordingly for non-zero porosity); a water ice projectile is assumed with $\rho_p = 0.91$ g cm$^{-3}$. {In all cases we assume $\nu = 0.4$ and $g=1.62$ \ms}
{The empirical parameter $K_1$ was measured for each target material in \S\ref{sec:crater}, and is typically of order unity \citep{Prieur2017}. Finally, we find $a = -0.890$ and $\mu' = 0.535$ reasonably describes all melt volume outcomes from our simulations (see \S\ref{sec:disc} for details). In this manner, we may investigate whether the simulation results agree with theoretical scaling relations for melt volume. Further, we may use the scaling relations to extend our analysis to a broader range of materials than those simulated, and investigate conditions most amenable to crater identification.
}

\setlength{\tabcolsep}{11.0pt}
\renewcommand{\arraystretch}{1.0}
\begin{table*}
\centering
\begin{tabular}{l c c l l r} 
 \hline
Case & $a$ & $\mu'$ & $K_1$ & $\mu$ & Case Description \\
 \hline\hline
S1 &  $-0.890$ & $0.535$ & 1.366   & 0.533  & Ice projectile, basalt target, $Y = 5$ Pa, $\Phi=0\%$ (this study) \\
S2 &  $-0.890$ & $0.535$ & 1.228   & 0.510  & Ice projectile, basalt target, $Y = 5$ Pa, $\Phi=12\%$ (this study) \\
S3 &  $-0.890$ & $0.535$ & 1.143   & 0.514  & Ice projectile, basalt target, $Y = 5$ Pa, $\Phi=20\%$ (this study) \\
S4 &  $-0.890$ & $0.535$ & 1.334   & 0.554  & Ice projectile, basalt target, $Y = 10^7$ Pa, $\Phi=0\%$ (this study) \\
S5 &  $-0.890$ & $0.535$ & 1.513   & 0.493  & Ice projectile, basalt target, $Y = 10^7$ Pa, $\Phi=12\%$ (this study) \\
S6 &  $-0.890$ & $0.535$ & 1.479   & 0.486  & Ice projectile, basalt target, $Y = 10^7$ Pa, $\Phi=20\%$ (this study) \\
\hline
E1 &  $-0.482^a$ & $0.624^a$ & 1.6   & 0.564  & Wet sand (proxy for competent rock)      \citep{Schmidt1987} \\
E2 &  $-0.482^a$ & $0.624^a$ & 1.4   & 0.381  & Dry quartz sand (proxy for porous rock)  \citep{Schmidt1987} \\
E3 &  $-0.482^a$ & $0.624^a$ & 1.615 & 0.558 & Basalt, wet sand analog,   $f=0.1$, $\Phi=0\%$ \citep{Prieur2017} \\
E4 &  $-0.482^a$ & $0.624^a$ & 1.585 & 0.516 & Basalt, porous sand analog $f=0.1$, $\Phi=12\%$ \citep{Prieur2017} \\
E5 &  $-0.482^a$ & $0.624^a$ & 1.984 & 0.394 & Basalt, porous sand analog $f=0.6$, $\Phi=12\%$ \citep{Prieur2017} \\
E6 &  $-0.482^a$ & $0.624^a$ & 1.473 & 0.424 & Basalt, porous sand analog $f=0.6$, $\Phi=40\%$ \citep{Prieur2017} \\
\hline
\end{tabular}
\caption{{Parameter combinations {for analytically linking melt volume, transient crater diameter, and impact velocity}. Columns correspond to case number ({S denotes simulated, E denotes extended}), melt volume scaling {constant and} exponent ($a$ and $\mu'$), {transient} crater diameter scaling coefficient ($K_1$), crater diameter scaling exponent ($\mu$), and a brief description of the case study. The combinations of parameters {are derived from our simulations in the top portion, and sample} different regimes reported by \citet{Schmidt1987} and \citet{Prieur2017} {in the bottom portion}. For specific scenarios from \citet{Prieur2017}, $f$ denotes coefficient of friction, and $\Phi$ denotes porosity.}
{$^a$For case studies not simulated in this study, we adopt identical $a$ and $\mu'$ from \citet{Barr2011}, which was found to hold for a variety of materials.}}
\label{tab:meltcombos}
\end{table*}

\begin{figure}
    \centering
    \includegraphics[width=\linewidth]{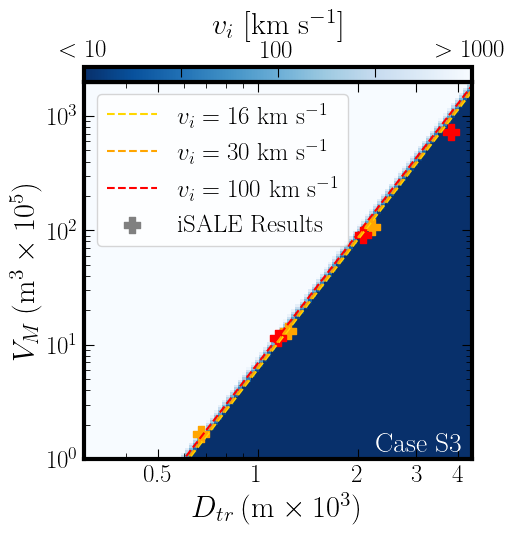}
    \caption{Analytic estimates of impact speed as a function of {melt volume} and diameter of transient crater {for Case S3 (see Table~\ref{tab:meltcombos}). The relationship between the three variables follows from scaling relations. Scaling parameters were fit using the simulated transient diameters and melt volumes. For reference, iso-velocity contours are plotted for 16 \kms\ (at which point the scaling relation becomes valid, and is also characteristic of asteroid impacts), 30 \kms, and 100 \kms, and the simulation results for 30 \kms\ and 100 \kms\ impacts are marked.}}
    \label{fig:melt}
\end{figure}

The relationship between $D_{tr}$, $V_M$, and $v_i$ is plotted in Figure~\ref{fig:melt} for {for targets with $\Phi=20\%$ in the gravity-dominated regime.} {The 10 \kms\ impacts are excluded, since the melt number is less than 30; the cutoff is at approximately 16 \kms. We plot contour lines for 16 \kms, as well 30 \kms\ and 100 \kms.} The {difference between the} diameter scaling exponent {$\mu$ and the melt volume scaling exponent $\mu'$} determines the {velocity spread across $D_{tr}$ and $V_M$}. Increasingly significant velocity dependence manifests as a more gradual gradient in the figure, and larger separation between constant velocity lines. {Results in Figure~\ref{fig:melt} are representative of the other porosities in that $\mu \simeq \mu'$, so the distance between velocity contours is small. This trend indicates that melt volume may not be a significantly differentiating metric for inferring projectile parameters, at least for the materials simulated here.}

{Nevertheless, other impact configurations may be more conducive for breaking degeneracy with combined $D_{tr}$ and $V_M$ measurements. The parametrization $a = -0.482$ and $\mu' = 0.624$ \citep{Barr2011} is suitable for impacts of identical target and projectile materials (spanning aluminum, iron, ice, dunite, and granite).}
We calculated melt volume for several parameter combinations that span the various regimes covered in {prior studies, as follows. The specific combinations are listed in Table~\ref{tab:meltcombos}}. Parameters from \citet{Schmidt1987} are empirical, where wet sand and dry sand were used as proxies for competent and porous rock, respectively. iSALE-2D simulations by \citet{Prieur2017} assumed a basalt target with variable coefficient of friction ($f$) and porosity ($\Phi$). \citet{Elbeshausen2009} simulated oblique impacts into granite with iSALE-3D, varying $f$ and $\theta$ with fixed $\Phi=0\%$. Since their coefficients are reported in terms of volume scaling ($\pi_V$), we do not consider specific instances of their simulations. They find ${\mu} \simeq {0.469}$ for $f = 0.7$ and ${\mu} = {0.548}$ for $f = 0.0$, which are comparable to some scenarios from \citet{Schmidt1987} and \citet{Prieur2017}.

\begin{figure*}
    \centering
    \includegraphics[width=\linewidth]{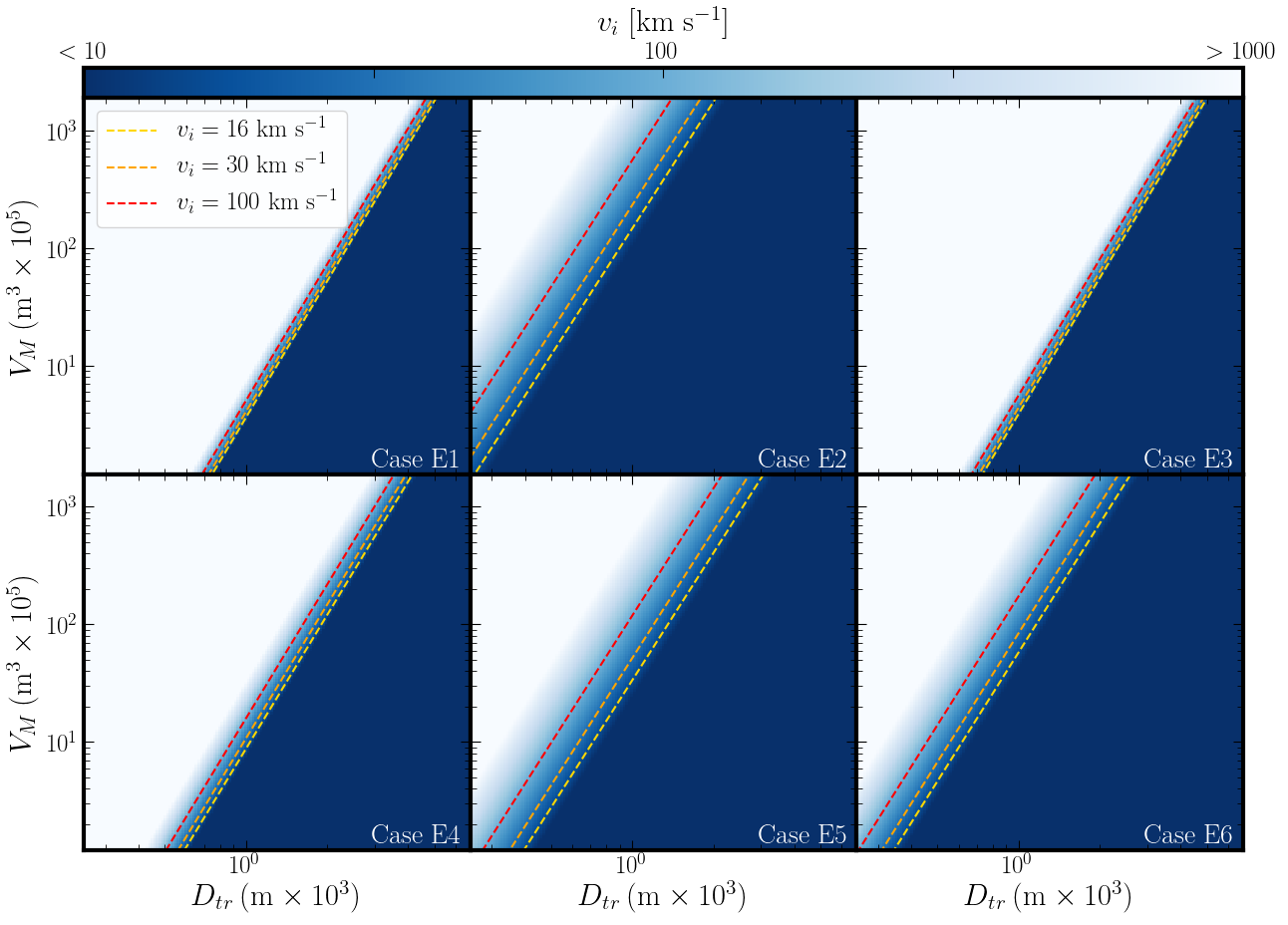}
    \caption{Analytic estimates of impact speed as a function of {melt volume} and diameter of transient crater, for six examples of target materials {that are explored theoretically without dedicated simulations in this study}. Panels show predictions for impacts {into targets with various scaling parametrizations (representing different materials), which are listed in Table~\ref{tab:meltcombos}.} Lines of constant {impact} velocity are drawn for {16 \kms, 30 \kms, and 100 \kms}, denoted by varying line colors. The larger the separation between these lines, the greater difference in melt volume produced in craters of the same diameter, and hence the easier to determine projectile characteristics.}
    \label{fig:meltExt}
\end{figure*}

{{Impacts} involving dry sand or porous basalt have the lowest values of {$\mu$}, and the melt volume for $v_i = 16-100$ \kms\ spans approximately {0.5 dex} at fixed transient crater diameter (Figure~\ref{fig:meltExt}).}  {In contrast to porous scenarios, wet sand results in the least spread, making it the most challenging for identifying ISO impact craters. We emphasize the critical importance that melt {approximately} scales {with energy for these materials}; else, velocity dependence effectively vanishes \citep{Abramov2012}. The results are encouraging for lunar melts that involve the unconsolidated regolith \citep{McKay1991} and lunar crust of porosity $\Phi \sim 10-20\%$ \citep{Kiefer2012}. In practice, $\mu$ and $K_1$ would both require tight constraints, and hence depend on whether the crater in question formed in the basaltic mare or anorthosite highlands. Additional considerations include impact angle and projectile density.}

\section{Discussion and Conclusions} \label{sec:disc}

{Intensive study of the lunar cratering record, including prospects for identifying ISO craters, will soon be forthcoming. {2020 marked the} first lunar sample return mission in nearly 45 years by the Chang'e 5 Lander \citep{Zeng2017}; this is a precursor to a modern-day surge in lunar exploration, as well as preliminary steps to establishing a permanent presence on Mars. We discuss how upcoming remote observations, return missions, and {\it in situ} analyses might assist in the identification of ISO impact craters.}

\subsection{Measuring Melt Volumes in Search of ISO Craters}

{In the previous section we showed that a high-speed ISO impact {can} yield a significantly enhanced melt volume {for certain projectile/target material combinations}. While other factors need to be accounted for, including impact angle and target material properties, melt volume can help break the degeneracy between impact velocity and projectile mass: specifically, by searching for craters that fall in the high melt volume, low diameter regime. {\it In situ} analyses \citep[e.g.][]{Grieve1982a} combine the percentage of melt in localized regions with crater geometry to obtain an overall estimate of melt volume. However, there are significant sources of uncertainty \citep{French1998}, such as strong dependence on target materials (e.g volatile content \citep[][]{Kieffer1980}), and modification processes \citep{Melosh1989}. Furthermore, detailed mapping of large melt volumes may be forbiddingly time- and resource-intensive for surveying candidate ISO craters, given there should be only of order unity high-speed ISO impact craters between the Moon and Mars (\S\ref{sec:vel}).}

{Remote sensing melt volume may be an appealing alternative to {\it in situ} analyses. Currently,} some of the best {remote-based} estimates rely on LROC images \citep{Plescia2014}, where melt pools are identifiable by low-albedo, flat crater floors. {Crater diameters may also be readily extracted from LROC images. The correspondence between final and transient crater diameters is non-trivial; however, a simple heuristic for would be to search for craters with particularly high ratios of melt volume to diameter as potentially of ISO origin.} {\citet{Plescia2014} estimate the melt volume by} fitting the crater wall profile, extrapolating the profile to depths below the melt pool, and taking the difference between the observed crater volume and that of the entire original crater. {They acknowledge the estimates are order-of-magnitude, since} additional melt may have been ejected from the crater, displaced onto the crater wall, or buried within the debris layer on the crater floor. {\citet{Silber2018} analyzed theoretical (from iSALE-2D) and observed \citep[][]{Plescia2014} melt volumes of lunar craters, with a similar goal as ours of breaking degeneracies between projectile characteristics. They were able to match the observed spread in melt volume ($\sim 2$ orders of magnitude) for a given crater diameter. Individual craters/projectiles were not investigated, and velocities only up to $20$ \kms\ were considered. {These results indicate that imaging may be a viable method of finding enhanced melt volumes --- however, given that remote sensing uncertainties are of the same order as the largest melt volume spreads for a fixed $D_{tr}$ (see \S\ref{sec:melt})}, higher precision followup measurements may be necessary; possibly {\it in situ}.}

{The precision in melt volume required to identify an ISO crater depends on target materials, apparent in the variable spread in Figure~\ref{fig:meltExt} panels. Lunar seismology (e.g. of small impacts) may soon be a feasible approach for estimating melt volumes without requiring assumptions of the subsurface crater geometry.} Arrival time anomalies of $p$ and $s$ waves are frequently used to map geological structures such as mantle plumes \citep{Nataf2000}, and are also employed for identifying and characterizing natural oil reserves. For our purposes, we note simple craters tend to have a `breccia lens' at their floors, which is a mixture of inclusion-poor breccia that formed immediately below the impact plus mixed breccia that formed due to the shear of melt sliding up the crater walls (this material collapsed during the modification stage) \citep{Grieve1987}. Appropriately placed sensors within and near {an existing} crater may allow seismic imaging of the breccia lens if the recrystallized melt has sufficiently different material properties from surrounding rock or there is a discontinuity in wave propagation between the crater wall and the breccia lens. {Seismic imaging of artificial shots/blasts has been applied extensively to the Chicxulub crater \citep{Gulick2013}, for example in identification of the top of its melt sheet \citep{Barton2010}. It was also used to measure melt volume in the Sudbury Basin \citep{Wu1995}.} {Seismic imaging could in principle extend to Moon for measuring melt volume; although it is still subject to uncertainties surrounding ejected or displaced melt during the crater's formation.}

\subsection{Petrological Considerations}

{In addition to producing more melt, faster impacts induce higher peak shock pressures. We discuss whether high-pressure petrology provides an alternative or complementary route to identifying ISO impact craters.
}

{Target material in our 100 \kms\ simulations} experienced higher peak pressures ($\sim 3$ TPa) compared to target material in the {30 \kms\ simulations (Figure~\ref{fig:profiles}). In both cases, the pressures are sufficiently high to produce {coesite}, stishovite and maskelynite \citep{Stoeffler1972, Melosh2007}, so high-pressure phases and polymorphs are probably insufficient criteria for identifying an ISO crater. However, the abundance or composition of vapor condensates might point to an ISO projectile}.
To date, only a handful of lunar vapor condensates have ever been found \citep{Keller1992, Warren2008}. {High-alumina, silica-poor material \citep[HASP,][]{Naney1976} deemed evaporation residue is complemented by {volatile-rich alumina-poor (VRAP) glasses and} gas-associated spheroidal precipitates (GASP). {These spherules are} attributed to liquid condensation droplets {(VRAP is enriched in volatiles like K$_2$O and Na$_2$O, whereas GASP is not; VRAP spherules are also about $200-400$ nm in diameter, whereas GASP spherules span roughly $2-10\,\mu$m).}} {VRAP/GASP are} identified by a distinct depletion of refractory species Al$_2$O$_3$ and CaO. The highest speed impacts (e.g. ISOs) may generate more {vapor} condensates, which may be detected in {surrounding} rock samples. Also, the  {exceptionally} high pressures generated in ISO impacts may alter the composition of residues and condensates; for example, pressures may be sufficient to shock vaporize Al$_2$O$_3$, CaO, or TiO$_{2}$, depleting them from HASP and enhancing them in {condensates}. Predicting the constituents of vapor condensates associated with ISO impacts will require mapping the high-pressure phase space for low-volatility target materials.

{Microscopic spherules were also produced through ancient lunar volcanism \citep{Reid1973}, but these spherules can be robustly distinguished from those of impact origin. \citet{Warren2008} used a combination of Al content, as well as trends of TiO$_2$ and MgO to establish an impact origin. As another example, \citet{Levine2005} ruled out a volcanic origin for $>90\%$ of 81 spherules in an Apollo 12 soil sample, based on low Mg/Al weight ratios. They also found a large fraction had $^{40}$Ar/$^{39}$Ar isochron ages younger than 500 Myr, which is inconsistent with known periods of lunar volcanism.}

Impact speed also influences the dimensions of vapor condensates. \citet{Johnson2012a} present a model for the {condensation} of spherules {from impact-generated rock vapor}. They find that the highest impact speeds yield smaller spherule diameters owing to higher speed expansion of the vapor plume {for impact speeds greater than $\sim 28$ \kms}. The vapor plume model of \citet{Johnson2012a} invokes a simplified plume geometry, and assumes the projectile and target are both comprised of SiO$_2$. {The same authors employed this model when they estimated projectile velocities and diameters for major impact events in Earth's history \citep{Johnson2012b}. In theory, particularly small} spherules may be linked to impact speeds consistent with ISOs. {\citet{Johnson2012a} explored} velocities up to 50 \kms, but extrapolation suggests 100 \kms\ impacts may produce spherules of diameter $\lesssim 10^{-7}$ m. Degeneracy with projectile size persists, but might be reconciled with, for example, crater scaling relationships. {In regards to identifying ISO craters on the Moon, a significant concern is that vapor condensate spherules may be scattered extremely far from the impact site. For example, microkrystite condensates from the K-T impact form a world-wide spherule layer \citep{Smit1992}. Isolating the crater of origin would likely require widespread mapping and classification of spherules on the Moon, which is beyond current capabilities.}

{Could melts or condensates be used to infer an ISO's composition? It is well understood that these impact products comprise a mixture of projectile and target material. For example, \citet{Smit1992} estimated that condensate spherules from the K-T impact contain a $\sim 10\%$ bolide component from their Ir content. Although, the task might be challenging for lunar vapor condensates because the spherules are microscopic and of extremely low abundance in the Moon's crust \citep[$<0.001\%$ by volume,][]{Warren2008b}. \citet{Keller1992} and \citet{Warren2008} do not make inferences regarding the composition of the projectile(s) that generated the VRAP/GASP spherules, and to our knowledge, there has not yet been any study that links these spherules or the HASP residue to a projectile's composition.}

{Encouragingly, a number of projectiles involved in terrestrial impacts have been geochemically characterized, primarily via rocks within and near the crater. \citet{Tagle2006} review major findings and methods. Elemental ratios of PGEs (Os, Ir, Ru, Pt, Rh, Pd), plus Ni and Cr, are particularly effective if multiple impactite samples are available, since then it is not necessary to correct for elemental abundances in the target. Isotope ratios $^{53}$Cr/$^{52}$Cr and $^{187}$Os/$^{188}$Os are also commonly employed. This precedent extends to lunar impacts, as \citet{Tagle2005} used PGE ratios in Apollo 17 samples to determine that the Serenitatis Basin projectile was an LL-ordinary chondrite. Since these methods are based on refractory species, ISOs may be difficult to characterize. 2I/Borisov contains a significant volatile component \citep{Bodewits2020} as most comets do, and volatiles would explain 'Oumuamua's anomalous acceleration \citep{Seligman2019}. If ISOs have a refractory component, then elemental and isotoptic ratios could separate them from other projectile classes and offer important insights into their composition.
} 

\subsection{{Influence of Impact Angle}}

{Crater dimensions are degenerate with impact angle, a parameter unexplored in this study. Indeed, the most probable impact angle of $45{^\circ}$ would yield a considerably different crater than a head-on collision, all other factors being equal. \citet{Davison2011} quantified how several crater properties depend on impact angle. For example, crater volume is approximately halved for a $45{^\circ}$ impact, but the crater remains symmetrical for impact angles $\theta$ greater than a threshold $\theta_e \sim 10-30{^\circ}$, depending on the target material. They also found crater depth scales with $\sin\theta$, and width with $\sin^{0.46}\theta$. Melt production exhibits strong dependence on impact angle, as shown by \citet{Pierazzo2000} through simulations of Chicxulub-type impacts. In their 20 \kms\ impact speed simulations, the volume of material shocked above 100 GPa at $\theta = 30{^\circ}$ was rougly half that of a head-on collision, and was trivial for $\theta < 15{^\circ}$. While melt volume scales with impact energy \citep{Bjorkman1987}, the scaling breaks down if only the vertical component is considered in oblique impacts (i.e. $(v_i\sin{\theta})^2$ \citep{Pierazzo2000}. Nevertheless, melt volume was found to be proportional to transient crater volume across variations in $\theta$, with oblique impacts producing asymmetric melts.}

{How crater properties change with joint variations in impact angle and impact speed, especially in $v_i > 100$ \kms\ regime, would be interesting for future investigation, albeit computationally expensive. The studies discussed above indicate that crater and melt asymmetries may prove useful for constraining angle of incidence. They also suggest the maximal pressures and melt volumes produced by real ISO impacts are probably lower than those attained in our simulations, and that real crater dimensions may exhibit different ratios than those of our simulated craters. The reduction in peak shock pressure may also eliminate certain pretrological indicators of a high-speed impact, such as vapor condensates.}

\subsection{{Analysis of Scaling Exponents}}

{In \S\ref{sec:crater}, we fit power law relationships to dimensionless parameters (Equations~\ref{eqn:pi1}-\ref{eqn:pi4}) to determine the transient diameter scaling exponent $\mu$. An accurate and precise $\mu$ is needed in order to gauge the efficacy of using melt volume to disentangle projectile properties (\S\ref{sec:melt}). Inspection of the top panel of Figure~\ref{fig:fits} (gravity regime) shows that data points for fixed velocity follow a local slope that deviates slightly from the global fitted slope. This effect is especially pronounced for the two porous scenarios. For example, in the $\Phi=20\%$ simulations, locally fitting a power law to outcomes of simulations with fixed projectile velocity yield $\mu$ ranging from 0.32 to 0.38 (increasing with decreasing impact velocity). This discrepancy from the global fit $\mu = 0.514$ may arise from an additional velocity dependence which is not incorporated into the Pi-group scaling framework. A similar anomaly was reported by \citet{Prieur2017} when comparing their results to those of \citet{Wunnemann2011}. Discrepancies in $\pi_D$ reached up to $10\%$ between the two sets of simulations, which were conducted at $12.7$ \kms\ and $5$ \kms, respectively. The premise that $\mu$ may depend on impact velocity has been noted before. For example, \citet[][]{Yamamoto2017} find dependence even when impact velocity greatly exceeds the target bulk sound speed. Their interpretation is that the dependency arises because the shock front pressure decays at a rate $q$, which itself depends on impact velocity. This dependency suggests that it may be necessary to run a grid of simulations, densely spanning $L$ and $v_i$ for a fixed target composition, in order to constrain allowed values for $\mu$.}

{We also comment on the melt volume scaling parameters $a$ and $\mu'$ in Equation~\ref{eqn:meltmu}. \citet{Barr2011} performed simulations of impacts involving identical projectile and target materials, and fit all outcomes simultanesouly to obtain $a = -0.482$ and $\mu' = 0.624$. Their simulations of an ice projectile striking dunite, which is the most similar scenario to our simulations, yielded $a = -1.78$ and $\mu' = 0.819$. This value for $\mu'$ is unexpectedly high since it exceeds the theoretical upper-bound of energy scaling. We attempted to independently determine these two parameters from melt volumes in our simulations. The 10 \kms\ impact velocity scenarios were excluded. While our fit involves only two velocities, all of the materials considered produce similar melt volumes for a given $L$ and $v_i$. Fitting all simulation outcomes simultaneously yielded $a = -0.890$ and $\mu' = 0.535$. The scaling exponent is considerably lower that that found by \citet{Barr2011}. It is closer to $\mu' = 0.432$ found by \citet{Pierazzo1997} for ice/ice impacts (although \citet{Barr2011} suggested this value was influenced by the choice of target temperature by \citet{Pierazzo1997}). The discrepancy between our result and that of \citet{Barr2011} could arise from our choice of basalt as a target material, or differences in the adopted EoS (Tillotson vs. ANEOS). We also evaluated $\mu'$ using the 80 CPPR simulations from Appendix~\ref{sec:appmeltvol} to make the fit robust against our melt volume correction scheme; however, we obtained a comparable $\mu' = 0.564$. To investigate $\mu'$ further, we ran additional $20$ \kms\ and $50$ \kms\ impact simulations for one configuration involving a cohesionless, porous target. Fitting all velocities $(20, 30, 50, 100$ \kms) yielded $\mu' = 0.584$, while fitting just the lowest two velocities yielded $\mu' = 0.623$. This finding suggests a possible breakdown of Equation~\ref{eqn:meltmu} for very high melt numbers for ice/basalt impacts. Still, though, $\mu' = 0.623$ remains significantly lower than the $\mu' = 0.819$ from \citet{Barr2011}. In \S\ref{sec:melt} we opt to use $a = -0.890$ and $\mu' = 0.535$. However, the uncertainty on $\mu'$ indicates that a dedicated investigation of melt volume scaling would be useful; specifically, ice projectiles under different EoS specifications, impacting various target materials at a range of velocities.}

\vfill\eject

\subsection{Conclusion}

In searching for craters produced by ISOs impacting terrestrial bodies, it is important to have a set of criteria that differentiate these craters from those produced by asteroids and comets. By analyzing local stellar kinematics, we show ISOs {that encounter} the Solar System at speeds of $\geq$ 100 \kms\ {impact the Moon and Mars at rates of $\sim 0.09$ per Gyr and $\sim 0.29$ per Gyr, respectively. Importantly, 100 \kms\ exceeds} the impact speeds of most small Solar System objects. Therefore, crater properties that depend strongly on impact speed may be especially pertinent. {Transient} crater {dimensions are} expected to obey late-stage equivalence. We compare two hydro simulations to show that it is difficult to distinguish simple craters formed by high- and low-speed impacts. Melt volume, on the other hand, {does not} follow the point-source limit \citep{Pierazzo1997}, and offers a possible avenue for identifying high-speed craters. This approach requires overcoming degeneracies with impact angle and target composition, and obtaining precise estimates of the melt volume. Alternatively, vapor condensate composition and spherule dimensions could be revealing of extremely fast impacts. Facilitated by upcoming crewed and robotic Moon missions, identifying ISO craters may soon be feasible through {\it in situ} or return analyses of impact crater samples.

\acknowledgements

We gratefully acknowledge the developers of iSALE-2D, including Gareth Collins, Kai Wünnemann, Dirk Elbeshausen, Tom Davison, Boris Ivanov and Jay Melosh. We acknowledge generous support from the Heising-Simons Foundation through Grant \#2021-2802 to Yale
University.

\bibliographystyle{aasjournal}
\bibliography{main}

\appendix

\section{Diameter Scaling Validation for Basalt/Basalt Impacts} \label{sec:appscale}

\begin{figure*}
    \centering
    \includegraphics[width=\linewidth]{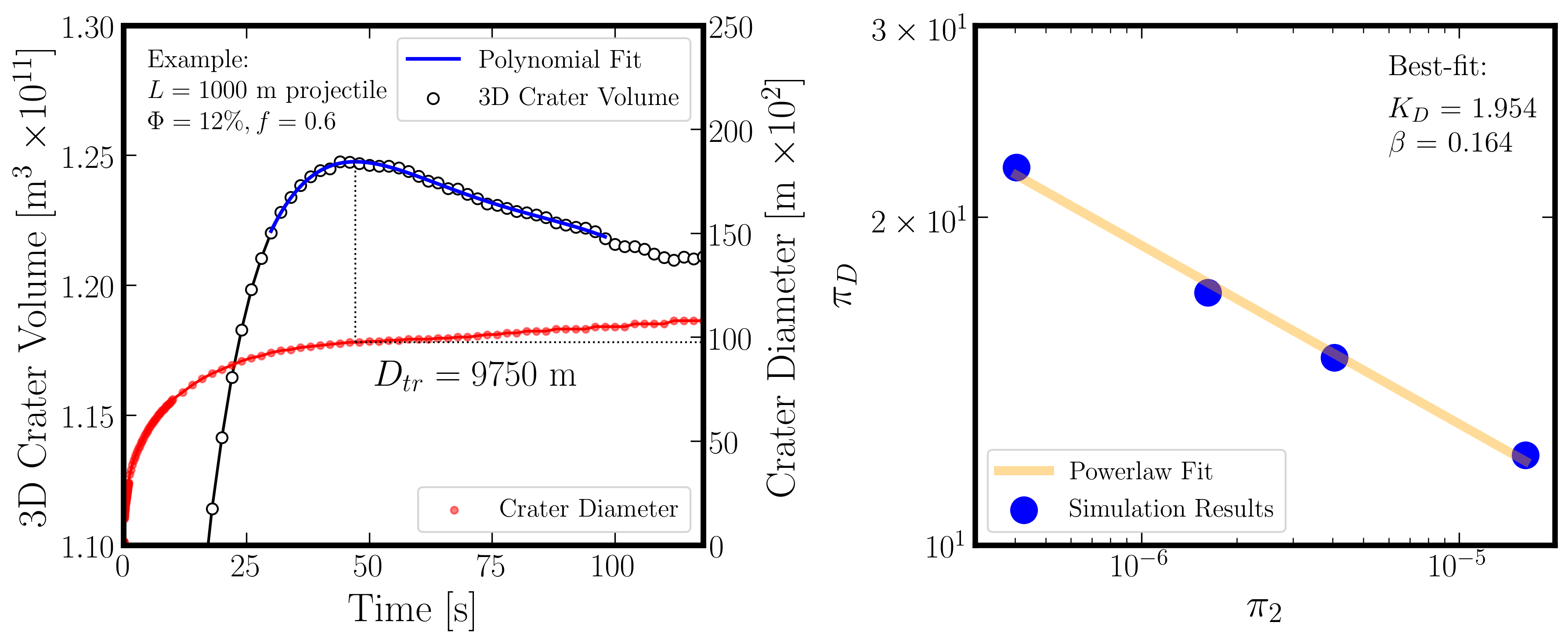}
    \caption{{Crater scaling relation for basalt-on-basalt impacts, which serves as verification of our simulation setup. {\it Left}: Determination of transient crater diameter for the simulation involving a 1000 m diameter projectile. A 5$^{\rm th}$-degree polynomial is fit to the crater dimensions near maximum volume. The diameter that maximizes the polynomial is taken as $D_{tr}$. {\it Right}: Best-fit power law plotted against dimensionless $\pi$-scaling quantities for the four simulations considered.}}
    \label{fig:verify}
\end{figure*}

{We perform additional simulations of a basalt projectile impacting a nearly cohesionless basalt target with $\Phi=12\%$ and $f=0.6$. These simulations serve as a foil to the ice projectile, and allow us to verify our simulation setup by independently measuring the associated transient diameter scaling relation, which was also measured by \citet{Prieur2017}. Since these craters are in the gravity-dominated regime, Equation~\ref{eqn:piD1grav} dictates the transient crater diameter. Impact speed was held constant at $v_i=12.7$ \kms, while projectile diameter took values of $L=25,100,250,1000$ m. \citet{Prieur2017} define $D_{tr}$ as the crater diameter at the time of maximum crater volume. In our simulations, crater volume as a function of time made discrete jumps as the crater grew in the extension zones. In order to make our measurement more robust to the spatial resolution, we fit a 5$^{\rm th}$-degree polynomial to volume as a function of time near its maximum value. We took the time at which the polynomial reaches its maximum as defining the transient crater. Finally, we linearly interpolated crater diameter between neighboring timestamps to obtain $D_{tr}$. Crater diameter was measured at the level of the pre-impact surface. The polynomial fit only included points within $2.5\%$ of the maximum crater volume, which helped exclude late-time data where the crater volume changes due to modification processes. Figure~\ref{fig:verify} shows a schematic of this process (left panel), and our results (right panel).}

{Our best-fit parameters are $K_D = 1.954$ and $\beta=0.164$, which agree well with the scaling from \citet{Prieur2017}, $K_D = 1.984$ and $\beta=0.165$. Values for $\pi_D$ predicted by our scaling relation and by that of \citet{Prieur2017} disagree by $<3\%$ for each of the projectiles we considered; this slight disagreement may be due to different zoning and resolution schemes in the simulation setups (e.g. we use a smaller high-resolution zone than \citet{Prieur2017}, the layer assigned to zero depth may be different, and cell sizes in the extension zone may also be different).
}

{Simulations in the strength-dominated regime (high target cohesion) required a different approach for measuring $D_{tr}$. In these cases crater volume grew during excavation and then plateaued, as opposed to reaching a maximum and subsequently decreasing. Crater diameter followed a similar trend. For these simulations, we select all timestamps in which the crater diameter is within $10\%$ of its diameter at the last simulation timestamp, and subsequently take the median of these diameters as a measurement of $D_{tr}$.}

\section{Melt Volume Dependency on CPPR}
\label{sec:appmelt}

\begin{figure*}
    \centering
    \includegraphics[width=\linewidth]{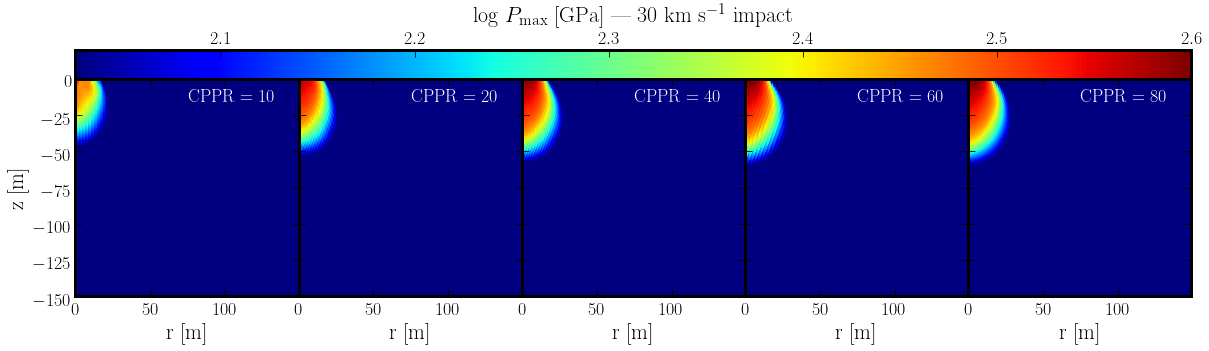}
    \includegraphics[width=\linewidth]{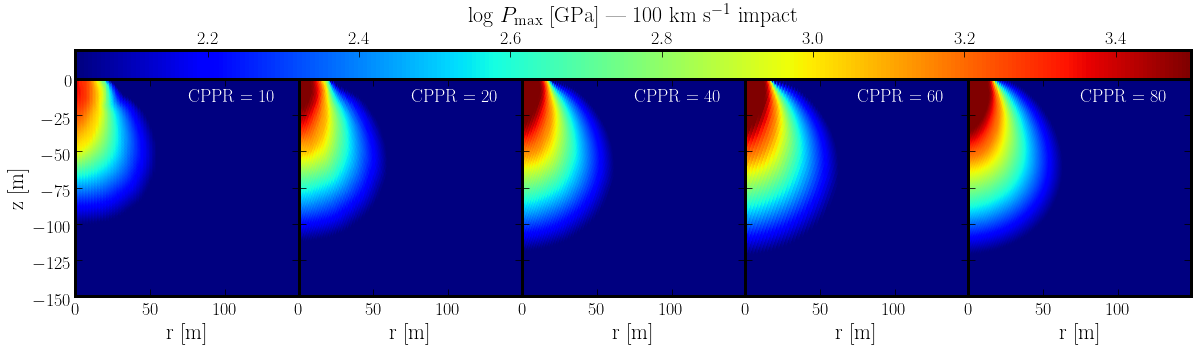}
    \caption{{
        Dependence of maximum pressure on simulation resolution for the nominal simulation setup in this study (ice projectile striking a cohesionless basalt target). Each panel depicts the original locations of tracer particles embedded in the high-resolution zone of the simulation, color coded by the maximum pressure they experience following the impact. The top row shows results for a 30 \kms\ impact, whereas the bottom row is for a 100 \kms\ impact. CPPR takes values of 10, 20, 40, 60, and 80. Note, the color scales for the top and bottom rows have different maxima, but the same minima at 106 GPa \citep{Quintana2015}.
    }}
    \label{fig:cpprmelt}
\end{figure*}

{Simulations in this study were conducted at a resolution of 20 CPPR, which was a compromise between simulation runtime and accuracy. \citet{Barr2011} found that near 20 CPPR, dunite/dunite impacts at 20 \kms\ underestimate melt volume by $\sim 15\%$. Since this study concerns different materials and impact speeds, we performed additional simulations to determine melt volume's dependence on CPPR. We simulated 40 m ice projectiles impacting (cohesionless) basalt targets at 10 \kms, 30 \kms, and 100 \kms\ at five different resolutions. Our results are depicted in Figure~\ref{fig:cpprmelt}. The volume of melt (plus vapor) was subsequently determined using the basalt complete melting pressure $P_c = 106$ GPa \citep{Quintana2015}. In the 30 \kms\ scenario, melt volume is underestimated by $45\%$, $22\%$, $8.3\%$, and $2.9\%$ for 10, 20, 40, and 60, respectively (compared to 80 CPPR). In the 100 \kms\ scenario, melt volume is underestimated by $38\%$, $19\%$, $7\%$, and $2.5\%$ for 10, 20, 40, and 60, respectively (again, compared to 80 CPPR). We found the 10 \kms\ impact simulations do not produce any melt at 80 CPPR or lower resolution. In our main analysis, we multiply melt volume by 1.28 and 1.23 in the 30 \kms\ and 100 \kms\ scenarios, respectively, to account for the resolution dependence.
}

\section{Analytic Melt Volume Scaling}
\label{sec:appmeltvol}

{As follows, we combine Pi-group scaling for a crater's transient diameter (Equation~\ref{eqn:piD1}) with a melt volume scaling relation (Equation~\ref{eqn:meltmu}) valid for $v_i^2/E_M \gtrsim 30$ \citep{Pierazzo1997}. Together, they break the degeneracy between projectile impact velocity and projectile size. While a useful demonstration, this analysis neglects impact angle dependence and detailed target lithology.}
For simple craters formed in granular targets, {Equation~\ref{eqn:piD1}'s} dependence on $\pi_3$ is negligible, and the relation follows:

\begin{equation} \label{eqn:piD1grav}
    \pi_D = K_1\pi_2^{-\mu/(2+\mu)}\pi_4^{(2+\mu-6\nu)/(6 + 3\mu)},
\end{equation}
for an empirically determined constant $K_1$. {The variables $\mu$ and $\nu$ follow from the point-source, coupling constant in Equation~\ref{eqn:couple}. They are often determined experimentally, and $\mu$ typically lies between} energy and momentum scaling ($\mu = 2/3$ and $\mu = 1/3$). {Let $\beta\equiv\mu/(2+\mu)$ and $\eta\equiv(2+\mu-6\nu)/(6 + 3\mu)$ to simplify notation.} {Next, consider the} scaling relation used by \citet{cintala1998}, which follows as a restatement of Equation~\ref{eqn:piD1grav} and multiplication of each side by the projectile volume to the one-third:

\begin{equation} \label{eqn:dtrscale}
    D_{tr} = \frac{K_1}{2} \Big(\frac{4\pi}{3}\Big)^{(1-\beta)/3} \Big(\frac{\rho_t}{\rho_p}\Big)^{\eta - 1/3} L^{1-\beta}g^{-\beta}{v_i^{2\beta}}.
\end{equation}
After calculating melt volumes from hydrocode simulations, they found a {power law} relationship for melt volume that depends strongly on $D_{tr}$ and weakly on $v_i$. 
By assuming the projectile is spherical, one may restate {Equation~\ref{eqn:meltmu}} as

\begin{equation} \label{eqn:vmscale}
    V_M = \frac{k\pi}{6}L^3\Big(\frac{v_i^2}{E_M}\Big)^{3\mu'/2}.
\end{equation}
Finally, combining Equations~\ref{eqn:dtrscale} and \ref{eqn:vmscale}, and in the process removing $L$ dependence, we arrive at

\begin{equation} \label{eqn:fullscale}
{
     V_M = \frac{k}{8}\Big(\frac{K_1}{2}\Big)^{-3/(1-\beta)}D_{tr}^{3/(1-\beta)}\Big(\frac{\rho_t}{\rho_p}\Big)^{(1-3\eta)/(1-\beta)}
     E_M^{-3\mu'/2} v_i^{3\mu'-6\beta/(1-\beta)}g^{3\beta/(1-\beta)}.
     }
\end{equation}
{
Similarly, we can derive a relationship between melt volume, transient crater diameter, and impact velocity in the case of strength-dominated craters. We start with}

{
\begin{equation} \label{eqn:piD1str}
    \pi_D = K_1K_2^{-\mu/2}\pi_3^{-\mu/2}\pi_4^{(1-3\nu)/3}.
\end{equation}
This equation involves a separate, empirically determined constant ($K_2$) and the same scaling variables as above. To simplify notation, let $\alpha \equiv \mu/2$ and $\xi \equiv (1-3\nu)/3$. Then,
}

{
\begin{equation} \label{eqn:dtrscale2}
    D_{tr} = \frac{K_1}{2} \Big(\frac{4\pi}{3}\Big)^{1/3} \Big(\frac{\rho_t}{\rho_p}\Big)^{\xi-1/3}\Big(\frac{\rho_t}{K_2Y}\Big)^{\alpha} L {v_i^{2\alpha}}.
\end{equation}
Combining the above equation with Equation~\ref{eqn:vmscale} yields
}

{
\begin{equation} \label{eqn:fullscale2}
     V_M = \frac{k}{8}\Big(\frac{K_1}{2}\Big)^{-3}D_{tr}^{3}\Big(\frac{\rho_t}{\rho_p}\Big)^{-3\xi+1}\Big(\frac{K_2Y}{\rho_t}\Big)^{3\alpha}
     E_M^{-3\mu'/2} v_i^{3\mu'-6\alpha}.
\end{equation}
Equations~\ref{eqn:fullscale} and~\ref{eqn:fullscale2} describe the theoretical amount of melt volume in gravity- and strength-dominated craters, respectively, accounting for different projectile and target bulk densities. Again, we emphasize that impact angle and lithology other than bulk density could influence the actual melt volume. Nevertheless, these equations give the baseline feasibility of determining a projectile's impact speed from measurements of its crater.}

\end{document}